\documentclass[iop]{emulateapj}

\newcommand{\etal}{\mbox{et~al.}}
\usepackage{multirow}
\usepackage{longtable}
\usepackage{url}

\shorttitle{Identification of sub-mJy radio sources in the E-CDFS}
\shortauthors{M. Bonzini et al.}

\begin{document}

\title{The sub-mJy radio population of the E-CDFS: optical and infrared counterpart identification.}


\author{M. Bonzini\altaffilmark{1},\email{mbonzini@eso.org}
V. Mainieri\altaffilmark{1},
P. Padovani\altaffilmark{1},
K.I. Kellermann\altaffilmark{2},
N. Miller\altaffilmark{3},
P. Rosati\altaffilmark{1},
P. Tozzi\altaffilmark{4},
S. Vattakunnel\altaffilmark{5},
I. Balestra\altaffilmark{4,6},
W. N. Brandt\altaffilmark{7},
B. Luo\altaffilmark{7},
Y. Q. Xue\altaffilmark{8,7}}


\altaffiltext{1}{ESO, Karl-Schwarzschild-Strasse 2, D--85748 Garching, Germany}
\altaffiltext{2}{National Radio Astronomy Observatory, 520 Edgemont
  Road, Charlottesville, VA 22903-2475, USA.}
\altaffiltext{3}{Department of Astronomy, University of Maryland, College Park, MD 20742-2421, USA}
\altaffiltext{4}{INAF Osservatorio Astronomico di Trieste, via G.B.
  Tiepolo 11, I-34131, Trieste, Italy}
\altaffiltext{5}{Dipartimento di Fisica Università di Trieste, piazzale Europa 1, I-34127 Trieste, Italy}
\altaffiltext{6}{Max Planck Institut f\"ur Extraterrestrische Physik,
  Giessenbachstrasse 1, D--85748 Garching, Germany}
\altaffiltext{7}{Department of Astronomy and Astrophysics, Pennsylvania State University,
University Park, PA 16802, USA}
\altaffiltext{8} {Key Laboratory for Research in Galaxies and
Cosmology, Department of Astronomy, University of Science and Technology
of China, Chinese Academy of Sciences, Hefei, Anhui 230026, China}



\begin{abstract}
We study a sample
of 883 sources detected in a deep Very Large Array survey at 1.4 GHz in the
Extended Chandra Deep Field South. The paper focuses on the
identification of their optical and infrared (IR) counterparts. We use
a likelihood ratio technique that is particularly useful when dealing
with deep optical images to minimize the number of spurious
associations. We find a reliable counterpart for 95\% of our radio
sources. Most of the counterparts (74\%) are detected at optical
wavelengths, but there is a significant fraction (21\%) only
detectable in the IR.  
Combining newly acquired optical spectra with data from the
literature we are able to assign a redshift
to 81\% of the identified radio sources (37\% spectroscopic). 
We also investigate the X-ray properties of the radio sources using the
\textit{Chandra} 4 Ms and 250 ks observations. In particular, we use a
stacking technique to derive the average properties of radio objects
undetected in the \textit{Chandra} images. The results of our
analysis are collected in a new catalog containing the position of the
optical/IR counterpart, the redshift information and the X-ray fluxes. 
It is the deepest multi-wavelength catalog of radio sources, which will 
be used for future study of this galaxy population.   
\end{abstract}



\section{Introduction}

Deep radio observations provide a powerful opportunity to investigate
the high redshift Universe. Moreover, since radio observations are
almost unaffected by dust extinction, they allow us to observe 
objects, which are heavily obscured in other bands. 
While bright radio sources are mostly powerful radio galaxies and 
radio-loud (RL) active galactic nuclei (AGN), at lower flux densities we 
observe an increasing fraction of star-forming galaxies (SFG) and
radio-quiet (RQ) AGN \citep[e.g.,][and references therein]{padovani09}. 

Source classification of deep radio surveys is not easy and requires
multi-frequency data. 
This approach was adopted in a series of papers, which studied a
radio selected sample of 266 objects in the Chandra Deep Field South
\citep[CDFS;][]{kellermann08,mainieri08,tozzi09,padovani09,
  padovani11}. Combining the information from different wavelengths,
these authors were able first to classify the sources as SFG, RQ AGN and RL AGN
and then to study the properties and the evolution of the different
classes separately. The promising results of this work have
encouraged us to apply it to a new radio catalog (N. Miller et al. 2012, in preparation)
that reaches a lower flux density limit and has a more
uniform coverage of the Extended-CDFS (E-CDFS).
As a consequence, we have three times more objects, with most of the
new sources in the sub-mJy regime (at 1.4 GHz). 

This paper focuses on the
identification of the optical and IR counterparts of the radio
sources. Our main goal is to assign a redshift to the radio sources
and to associate them with the correct photometry. This information will
be then used in future papers to classify the sources and study
their evolutionary properties. 
As mentioned above, 
a faint radio selected
sample includes sources of widely different nature: SFG with a blue stellar
population together with radio galaxies commonly hosted in redder
objects, and obscured AGN together with bright unobscured quasars.  For this
reason, it is important to consider a large wavelength range, from the
ultraviolet to the mid-infrared (MIR). 
Faint radio sources often correspond to faint
optical counterparts. Therefore, deep optical observations are
needed. This has an impact on the methodology that should be adopted
in the identification process. 

The structure of the paper is as follows: in Section \ref{sec-data} we present
the datasets, while in Section \ref{sec-count_ident-result} we describe the
likelihood ratio method we used to identify the counterparts of our sources 
and the results of the identification process, including an estimate of the spurious
association fraction and a comparison with the cross-correlation
method.  Section \ref{sec-redshift} discusses the redshift
distribution (spectroscopic and photometric) of our sample.  In
Section \ref{sec-X-ray} we deal with the X-ray counterparts of the radio
sources, while the description of the released catalog is given in Section
\ref{sec-final-cat}. In Section \ref{sec-discussion} we discuss our results and report our conclusions. Finally, in Appendix \ref{spectra_images} we present
new redshifts and spectra for the optical counterparts of 13 Very Large Array (VLA) sources and in Appendix \ref{sec-particular} we report on some peculiar sources. In this paper we use
magnitudes in the AB system, if not otherwise stated, and we assume a
cosmology with $H_{0}=70$ km s$^{-1}$ Mpc$^{-1}$, $\Omega_{M}=0.3$ and
$\Omega_{\Lambda}=0.7$.


\section{Data}
\label{sec-data}
\subsection{The radio catalog}

\begin{figure}
	\centering
	\includegraphics[width=\columnwidth]{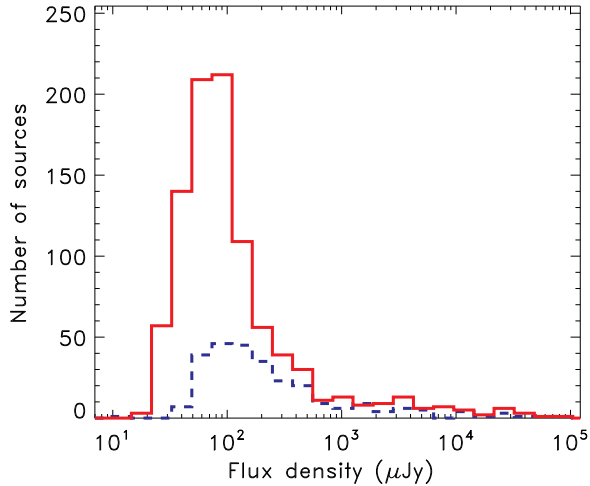}
 \caption{\small{Flux density distribution of the 5$\sigma$ E-CDFS radio catalog (solid line) compared to the sample described in \citet{kellermann08} (dashed line). The 5$\sigma$ sample is about three times larger and the majority of the sources have sub-mJy flux densities.}}
 \label{Si_distr}
\end{figure}

The E-CDFS was observed at 1.4 GHz with the VLA between 2007 June and September \citep{miller08}. The mosaic image covers an area of about $34\times 34$ arcmin with near-uniform sensitivity. The typical rms is 7.4 $\mu$Jy for a 2.8$\arcsec \times 1.6 \arcsec$ beam. 
The second data release (N. Miller et al. 2012, in preparation) provides a new source catalog with a 5$\sigma$ point-source detection limit, for a total of 883 sources.
We assigned a progressive identification number (RID) to the sources
ordered by increasing right ascension. 
The flux density distribution of the sample is shown in Fig. \ref{Si_distr}, where we use the peak flux density or the integrated flux density according to the specifications of N. Miller et al. (2012, in preparation). The median value of the distribution is
58.5 $\mu$Jy and the median signal to noise ratio (S/N) is 7.6. 
We note that $\sim 90\% $ of the sample has flux density below 1 mJy, a regime where RQ AGN and SFG become the dominant populations (e.g., \citealt{padovani09,padovani11}).
A classification of the radio sources will be presented in M. Bonzini et al. (2012, in preparation).

\subsection{Auxiliary data}
\label{sec-auxiliary}
\begin{figure*}
\centering
\begin{tabular}{c c c}
\includegraphics[width=0.65\columnwidth]{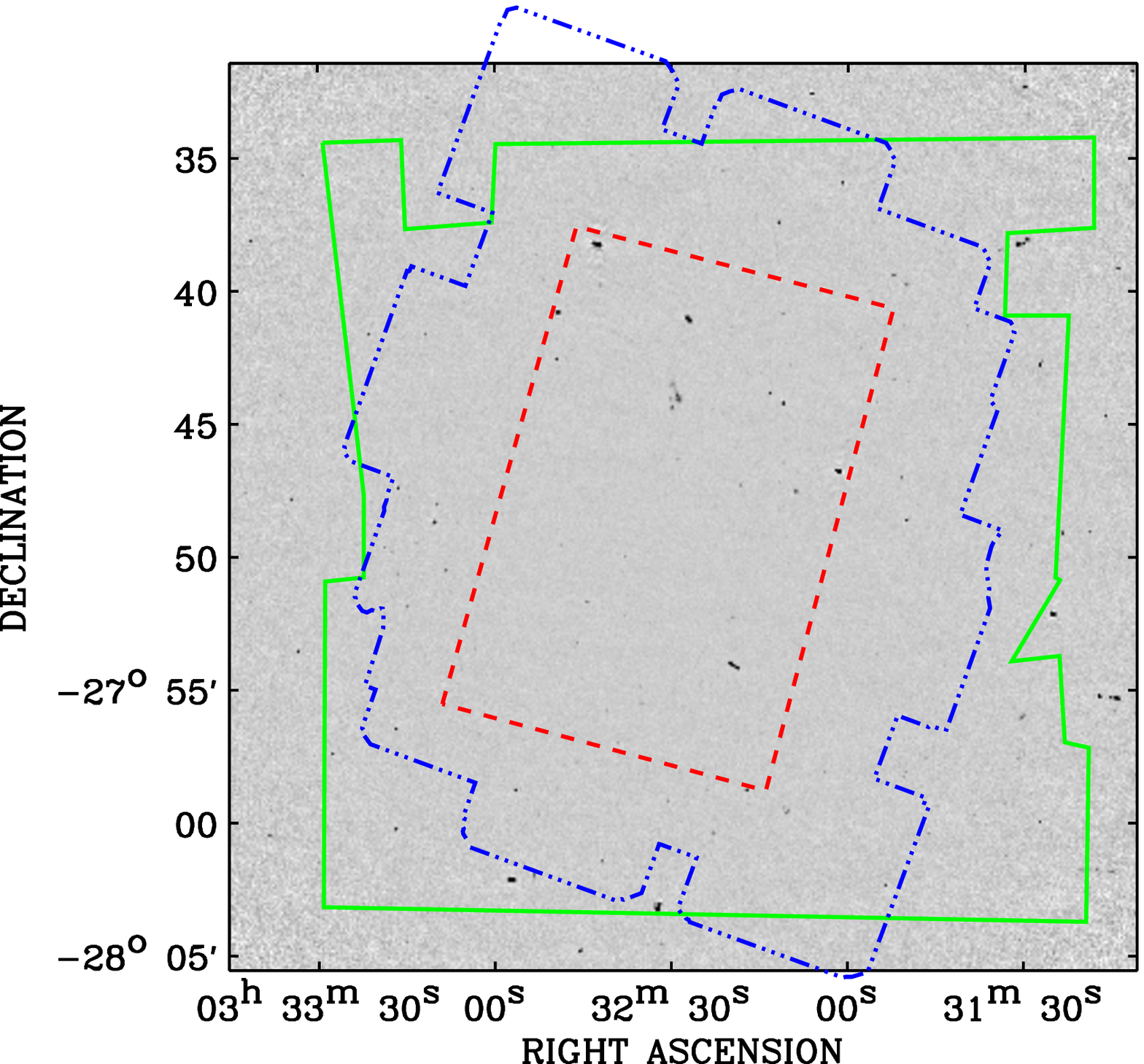} & \includegraphics[width=0.65\columnwidth]{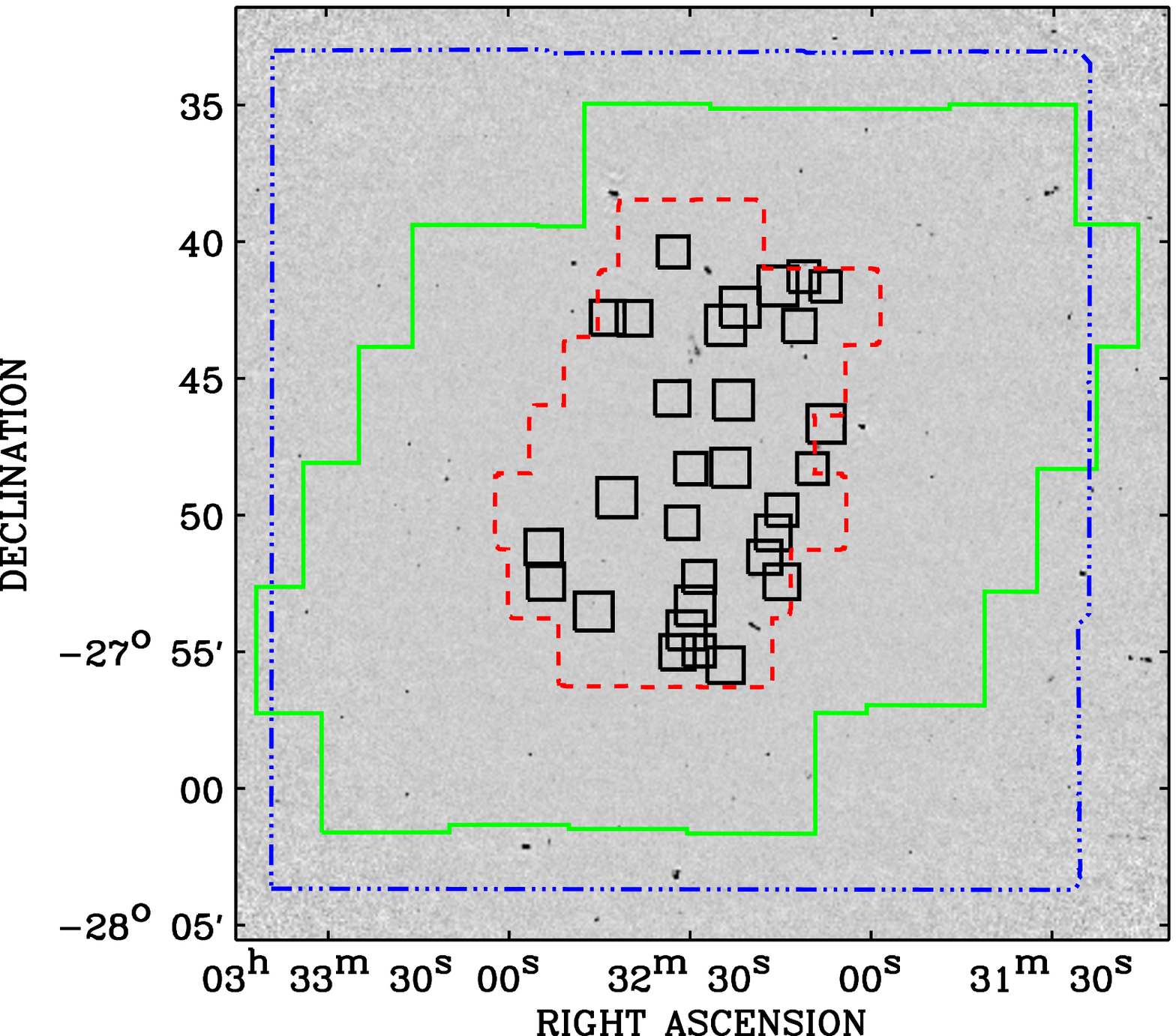} & \includegraphics[width=0.65\columnwidth]{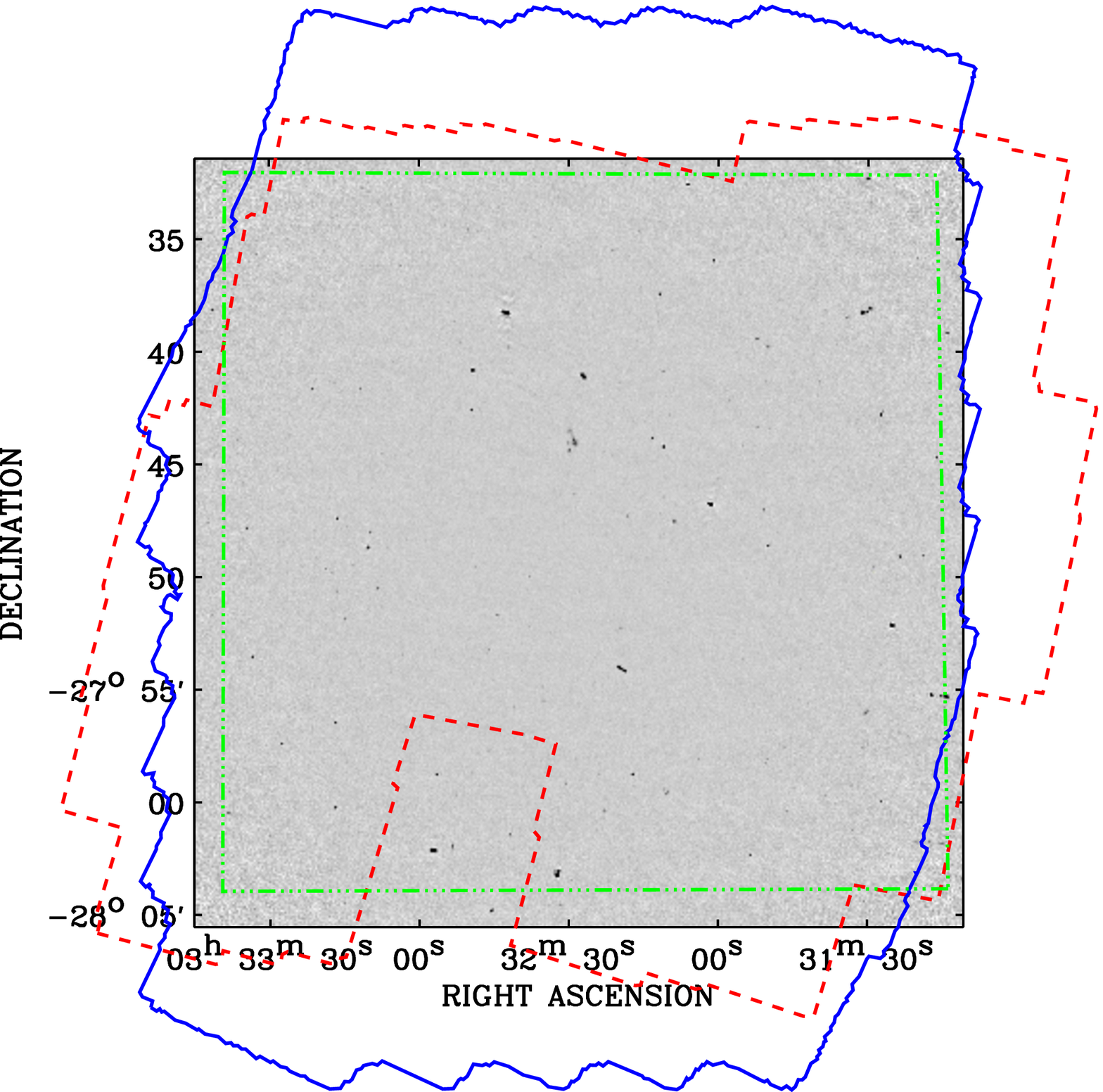} \\
\end{tabular}
\caption{\small{Multi-wavelength coverages overplotted on the VLA image. \textit{Left:} Optical catalogs (from outside): v-GEMS (solid line), U-VIMOS (dot-dashed line), and z-GOODS (dashed line). The R-WFI catalog coverage exceeds the VLA image and therefore is not plotted. \textit{Middle:} NIR catalogs (from outside): Ks-MUSYC (dot-dashed line), H-SOFI (solid line), Ks-ISAAC (dashed line) and H-GNS (small squares). \textit{Right:} MIR catalogs: IRAC-SIMPLE (dashed line) and 24um-FIDEL (solid line). The dot-dashed line encloses the area with photometric redshift catalogs coverage (see Section \ref{sec-redshift}).}}
\label{aux_cat_coverage}
\end{figure*}

The E-CDFS is one of the most studied patches of the sky and has been
observed in many wavebands. As we will discuss in the
following sections, this wealth of data is crucial to select the
correct counterpart of a radio source. Here we describe the large
amount of optical and IR data used in this work.  We considered a total of ten
catalogs. The complete list is reported in Table \ref{tab_aux_cat}
together with some basic information: the instrument used (column 2), 
the effective wavelength (column 3), the typical point spread function (PSF) 
(column 4), the 5$\sigma$ AB magnitude
limit (column 5), and the total area covered (column 6). The latter is
also shown in Fig. \ref{aux_cat_coverage}, where the footprint of each
mosaic image is plotted over the VLA image. For details on the
different data sets, we refer to the papers listed in column 7.
We divide the auxiliary catalogs in three groups according to their
selection band: optical, near-infrared (NIR) and mid-infrared
(MIR). The first group includes U-VIMOS, v-GEMS, R-WFI, and
z-GOODS. We note that the Wide Field Imager (WFI) observations are the only optical images
covering the whole VLA area. Therefore, even if they are shallower
than the others and with a lower spatial resolution, they were
crucial in the identification process. The U-VIMOS catalog has been produced by
us using the SExtractor software \citep{bertin96} from the original
images. In the NIR, we used the H-GNS,
H-SOFI, Ks-MUSYC and Ks-ISAAC catalogs. The H-GNS data, consisting of
30 pointed observations, cover a very small area of the E-CDFS but have a better resolution compared to the ground-based observations. At longer wavelengths,
the E-CDFS was mapped with the \textit{Spitzer Space Telescope} as
part of the SIMPLE and FIDEL surveys. These data are particularly
useful to identify high redshift sources (see Sec. \ref{sec-z-distr}
for details).

\begin{table*}
\footnotesize
       \caption{Auxiliary catalogs used for the identification of the radio sources counterparts. A description of the columns is given in Section \ref{sec-auxiliary}.}
\label{tab_aux_cat}
\begin{minipage}{0.9\textwidth}
\centering
\begin{tabular}{lccccccl}
\hline
\hline
 & (1)     & (2)        & (3)             & (4) 		      & (5)  & (6) 	    & (7)\\
 & Catalog$^a$ & Instrument & $\lambda_{eff}$ $^b$ & PSF FWHM &  AB mag  & Area     & References \\
 &	   &					& ($\mu$m)	  & 	  (arcsec)    &  (5$\sigma$ limit)  & (arcmin$^2$) &		\\
\hline
\noalign{\smallskip}
 \multirow{4}{*}{Optical} & U-VIMOS & VIMOS/ESO VLT & 0.390 & 0.2 & 28.0 & $\sim$800 & \citet{nonino09}\\
                          & v-GEMS  & ACS/\textit{HST}   & 0.578 & 0.1 & 28.5 & $\sim$800 & \citet{rix04}\\
                          &			&		&		&		&		&				&\citet{caldwell08}\\
                          & R-WFI   & WFI/ESO 2.2m  & 0.654 & 0.8 & 25.5 & $>$1100 & \citet{giavalisco04}\\
                          & z-GOODS & ACS/\textit{HST}   & 0.912 & 0.1 & 28.2 & $\sim$160 & \citet{dickinson03}\\
                          &			&		&		&		&		&				&\citet{giavalisco04}\\

\noalign{\smallskip}
\hline
\noalign{\smallskip}
 \multirow{4}{*}{NIR    } & H-GNS   & NICMOS/\textit{HST} & 1.607 & 0.2 & 26.5 & $\sim$43 & \citet{conselice11}\\
                          & H-SOFI  & SOFI/ESO NTT   & 1.636 & 0.55 & 22.0 & $\sim$800 & \citet{olsen06}\\
                          & Ks-ISAAC & ISAAC/ESO VLT & 2.745& 0.4 & 24.7 & $\sim$131 & \citet{retzlaff10}\\
                          & Ks-MUSYC & ISPI/CTIO 4m & 3.323& 0.3 & 22.3 & $\sim$900 & \citet{taylor09}\\
\noalign{\smallskip}
\hline
\noalign{\smallskip}
 \multirow{2}{*}{MIR    } & IRAC-SIMPLE & IRAC/\textit{Spitzer} & 3.507-4.436$^c$ & 1.7 & 23.8-23.6 & $>$1100 & \citet{damen11}\\
                          & 24um-FIDEL & MIPS/\textit{Spitzer} & 23.209 & 6 & 20.2 & $>$1100 & \citet{dickinson07}\\
\noalign{\smallskip}
\hline

\noalign{\smallskip}

\end{tabular}
\end{minipage}
$^a$ Name used in the text to refer to a specific catalog.\\
$^b$ Filter effective wavelength.\\
$^c$ The catalog is obtained from a combined image of 3.6$\mu$m and 4.5$\mu$m IRAC bands.\\

\end{table*}
\normalsize

\footnotesize
\normalsize

\section{Counterpart identification method}
\label{sec-count_ident-result}
\subsection{Likelihood ratio technique}
\label{sec-ident}
The first step in the identification process consisted of registering
each auxiliary catalog to the astrometric frame of the radio image, correcting for the median offsets between the radio
and the auxiliary catalogs. An average number of 400 sources was used to perform this registration and the typical median offset is 0.2$\arcsec$.
As already mentioned, a simple cross-correlation method, where the counterpart is selected as the closest object to the radio
source given a threshold matching radius, can lead to a
large number of spurious association when dealing with deep optical
images. Therefore, we adopted a likelihood ratio technique
\citep[e.g.,][]{sutherland92,ciliegi03}. This method allows us to take
into account not only the position of the counterpart, but also the
background source magnitude distribution and the presence of multiple possible
counterparts for the same radio source. Here we briefly describe this
technique following the formalism described in \citet{ciliegi03}. It
consists of three main steps:
\begin{enumerate}
\item[(a)] Compute the surface density of background sources $n(m)$ as a function of magnitude $m$.
\item[(b)] Evaluate the likelihood ratio ($LR$) for each possible
  counterpart.
\item[(c)] Compute the reliability ($rel$) of each association.
\end{enumerate}
(a) The magnitude distribution of background sources is obtained by 
counting all the sources within a 30$\arcsec$ radius around each radio object
and dividing them in magnitude bins ($\Delta m=0.5 $). The size of the
searching radius is set to contain, on average, just one radio source
and a substantial number ($>$100) of background sources for the deep
optical catalogs. The surface density of background objects $n(m)$ is
then computed dividing the obtained distribution by the total
searching area ($\pi \times (30\arcsec)^2\times N_{radio\ sources}$).

\begin{figure*}
\centering
\begin{tabular}{c c c}
\includegraphics[width=0.65\columnwidth]{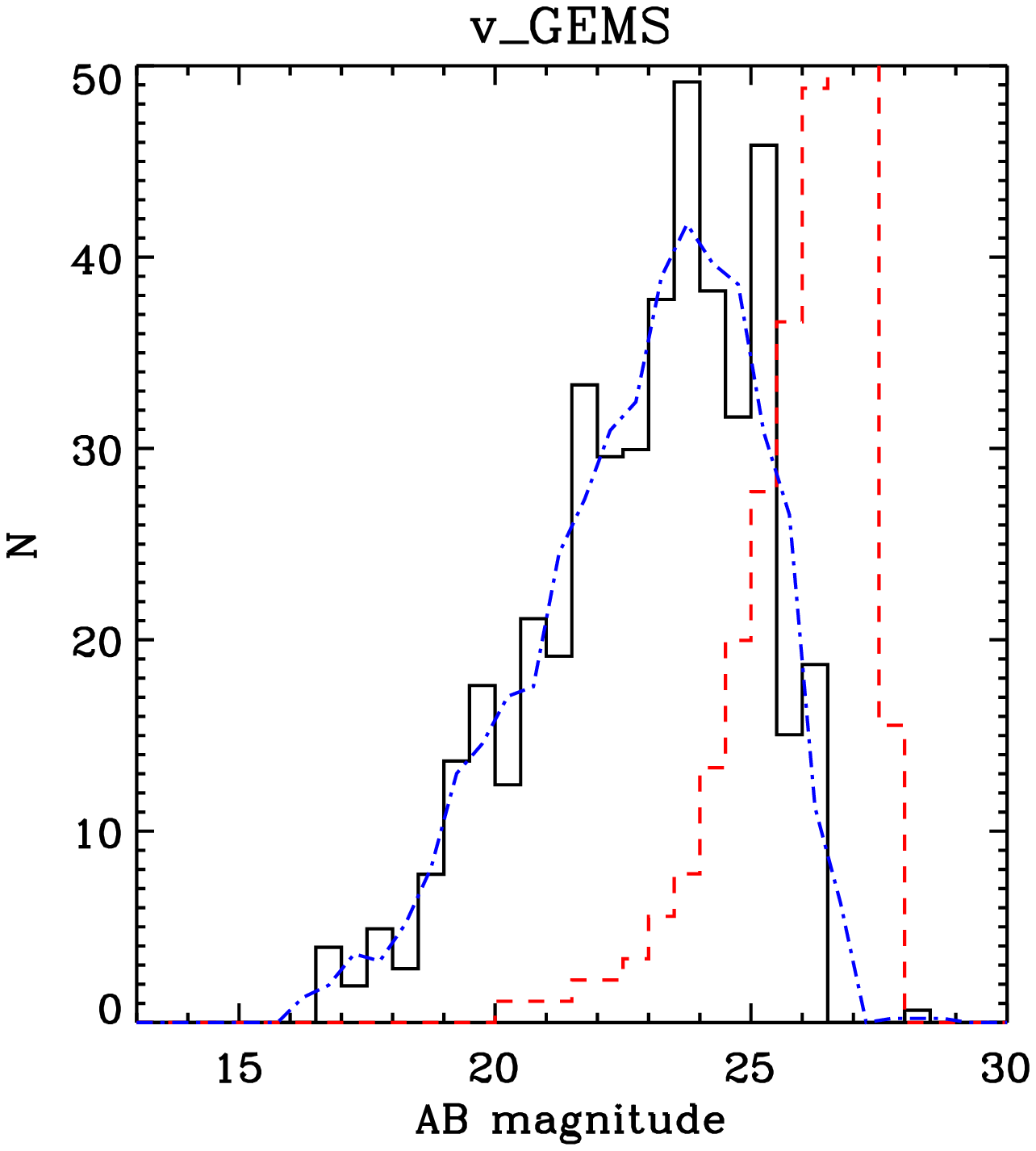} & \includegraphics[width=0.65\columnwidth]{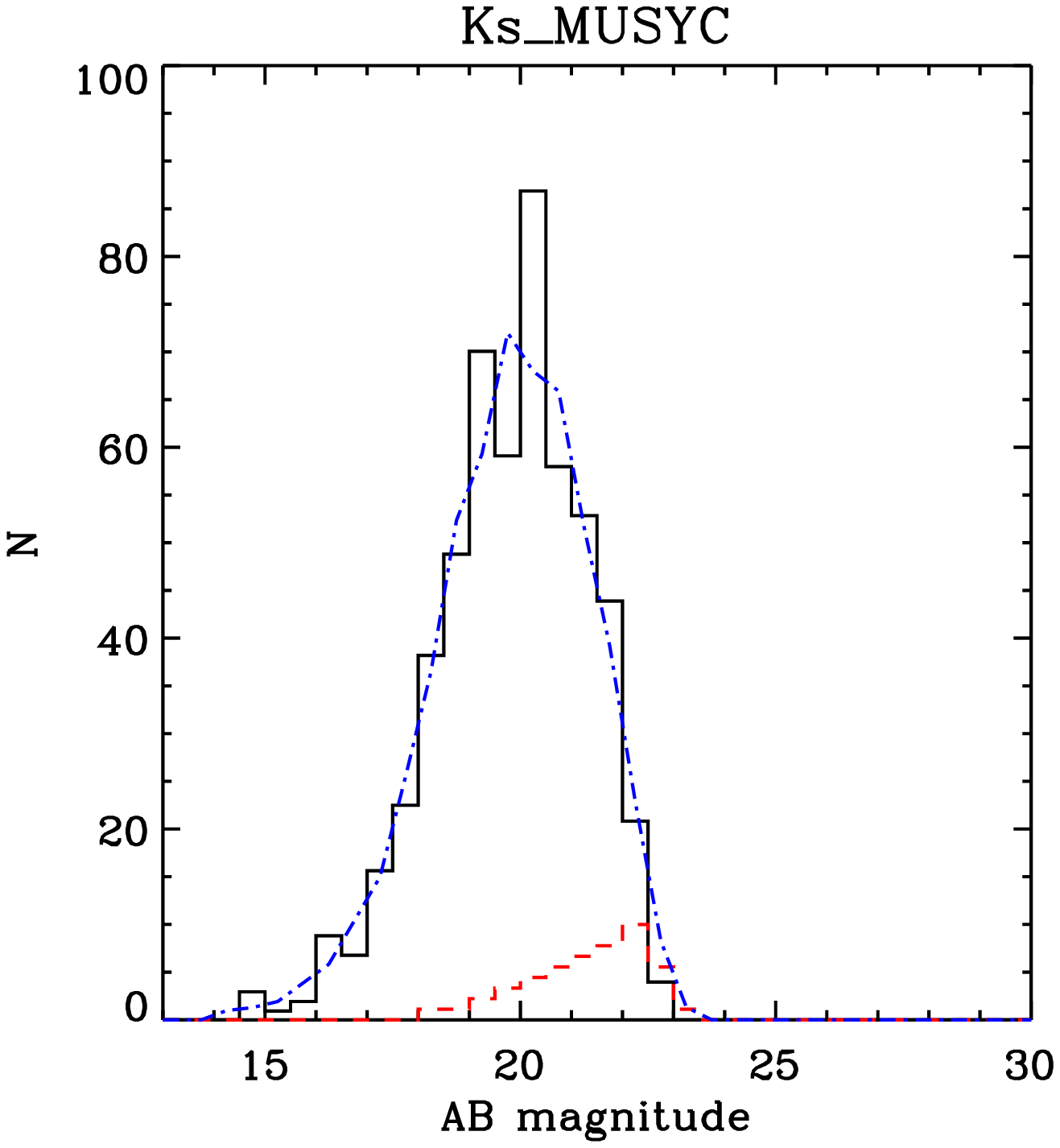} & \includegraphics[width=0.65\columnwidth]{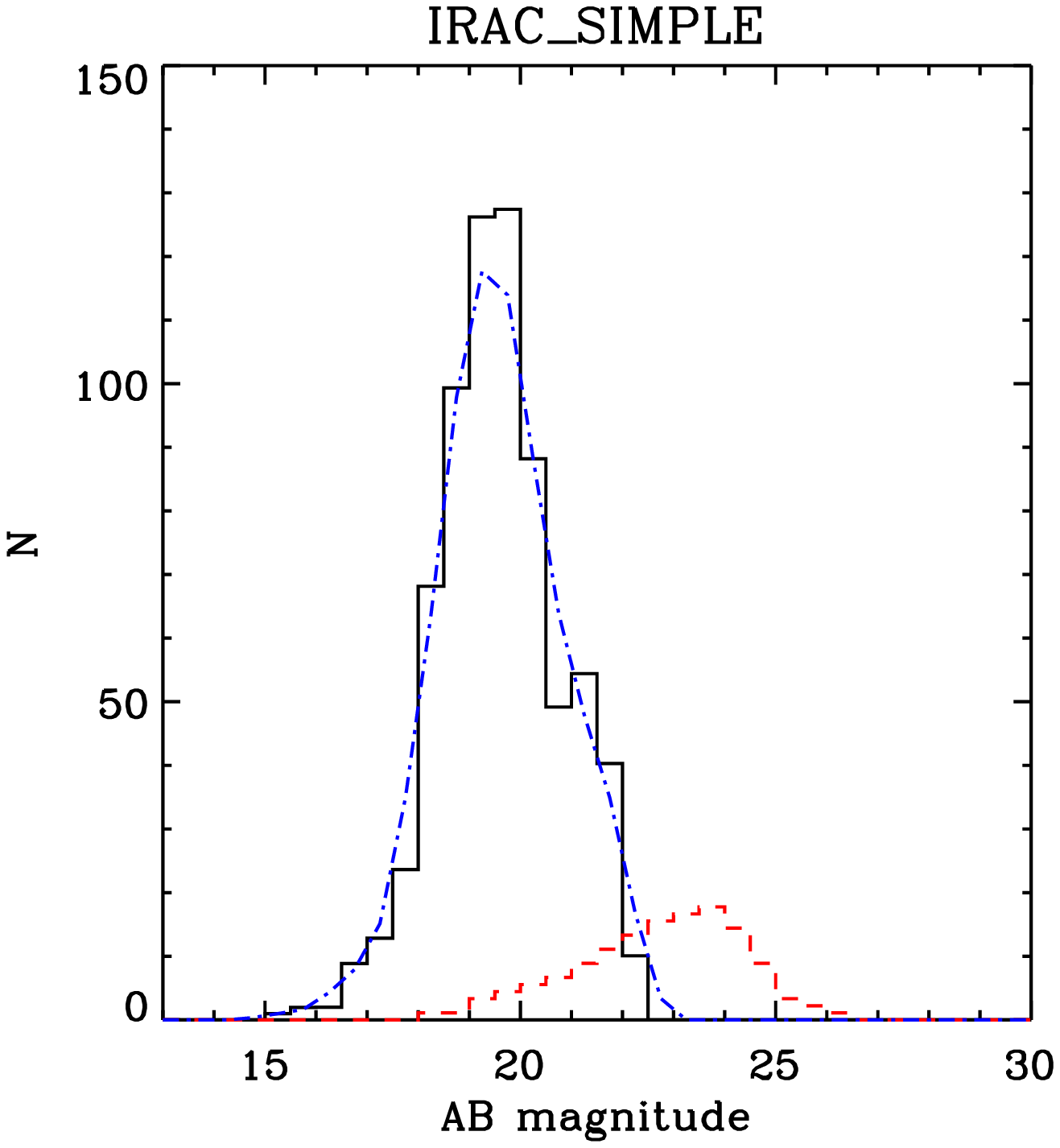} \\
\end{tabular}
\caption{\small{Background source distribution, $n(m)$, multiplied for the searching area ($\pi \times (2\arcsec)^2\times N_{radio\ sources}$) (dashed line) and the background subtracted distribution of counterparts, $real(m)$, (solid line) for the v-GEMS catalog (on the left), the Ks-MUSYC catalog (in the middle) and the IRAC-SIMPLE catalog (on the right). The dotted dashed line is obtained by smoothing the $real(m)$ distribution and it is used to compute $q(m)$.}}
\label{aux_cat_realm}
\end{figure*}

(b) The likelihood ratio for a counterpart candidate is defined as the
ratio between the probability that the source is the correct
identification and the corresponding probability for an unrelated
background source. 
Therefore we compute $LR$ as:
\begin{equation}
\label{equation_LR}
LR = \frac{q(m)f(r)}{n(m)}
\end{equation}
where $f(r)$ is the probability distribution function of the positional errors and $q(m)$ is the expected distribution of the counterparts as a function of $m$.
As for $f(r)$ we adopt a two-dimensional Gaussian distribution of the form:
\begin{equation}
\label{equation_fr}
f(r) = \frac{1}{2\pi \sigma}\exp{(\frac{-r^2}{2\sigma^2})}
\end{equation}
where $\sigma$ is the average between $\sigma_x = \sqrt{er^2_{aux} +
  er^2_{\alpha}}$ and $\sigma_y = \sqrt{er^2_{aux} +
  er^2_{\delta}}$. $\ er_{\alpha}$ and $er_{\delta}$ are the radio
positional errors given by the beam size ($1.6\times2.8$ arcsec)
divided by two times the S/N ratio of the considered source. To
account for further uncertainties on the VLA position, we added in
quadrature 0.1$\arcsec$ to the radio positional error (N. Miller et al., 2012, in preparation). The average positional error $er_{aux}$ for the optical catalog is $0.1\arcsec$ and 0.3$\arcsec$ for the others, with the exception of H-GNS
($er_{aux}=0.2\arcsec$) and 24$\mu$m-FIDEL ($er_{aux}=0.6\arcsec$) (see references given in Table \ref{tab_aux_cat}).
To derive an estimate for $q(m)$, we first counted all the objects in
the auxiliary catalog within a radius of 2$\arcsec$ around each radio
source. Then, we subtracted the distribution of background objects
computed on the same area ($n(m)\times \pi \times (2$"$)^2\times
N_{radio\ sources}$). The latter is shown in Fig. \ref{aux_cat_realm}
(dashed line), from left to right, for an optical, NIR and MIR
catalog, respectively. The background subtracted distribution,
$real(m)$, is plotted in the same figure as a solid line. 
Finally, we normalized the
distribution function as:
\begin{equation}
\label{formula_qm}
q(m) = \frac{real(m)}{\sum_i real(m)_i} \times Q
\end{equation}
where the sum runs over the total number of objects in the $real(m)$
and $Q$ represents the probability that the real counterpart is above
the catalog detection limit. As already verified by \citet{ciliegi03}
and \citet{mainieri08}, the number of identifications and the
associated reliabilities have a mild dependence on $Q$. Therefore, we
adopted a fixed value $Q=0.8$ for all the auxiliary catalogs as it corresponds to the average expected fraction of identifications.
Finally, combining $q(m)$, $f(r)$ and $n(m)$ according to eq. \ref{equation_LR}, we
computed the LR for each source in the auxiliary catalogs.

(c) The LR does not contain information about the possible presence of
many counterpart candidates in the surrounding of a specific radio
source.  It is therefore useful to define the reliability of each
association as:
\begin{equation}
rel_j = \frac{(LR)_j}{\sum_i(LR)_i + (1-Q)}
\end{equation}
where the sum is over all the candidate counterparts for the same
radio source \citep{sutherland92}.

\subsection{Identification results}
\label{ident_results}
Following the method described in the previous section, we built a
list of possible counterparts for each auxiliary catalog.  Initially,
we set a very low likelihood threshold ($10^{-6}$) to be sure not to
lose any counterpart. After a careful analysis, we decided to consider
as reliable only counterparts with reliability greater than 0.6. This
threshold ensures that the expected number of spurious associations is
below 5\% for each auxiliary catalog (see Section \ref{sec-surious}),
and at the same time maximizes the number of identified sources. The
identification rate for each auxiliary catalog is reported in column
(3) of Table \ref{tab-stat_each_cat}. The number of identified sources
is weighted by the number of radio sources inside the area covered
by each survey reported in column (2). We note that the number of
identifications increases with wavelength, from $\sim 65\%$ in the
optical catalogs up to $87\%$ in the MIR. That means that most of the
radio sources have a counterpart candidate in more than one auxiliary
catalog, and that there is a fraction of sources that are not
detected in the optical but only in the IR. In more detail, there are
652 radio sources (74\%) that have a counterpart in at least one of
the four optical catalogs, 76 (9\%) that have no counterpart in any of
the optical catalogs, but that are identified in at least one of the
NIR catalogs, and 111 (12\%) 
that have a counterpart only in the MIR. 
We will refer to these three groups as
optical, NIR and MIR selected counterparts, respectively. We
anticipate that they have different redshift distributions, with
NIR and MIR selected sources having on average higher redshift (see Section
\ref{sec-z-distr}). High redshift objects, thanks to their positive
K-correction, are more easily observed in
the IR than in the optical and this explains the higher identification
rate observed in the MIR catalogs.

\begin{table*}
     \caption{\small{Counterparts identified in each catalog and spurious association estimate for both the likelihood and cross-correlation methods. In column 2 we report the number of radio sources inside the area of the survey (see Fig. \ref{aux_cat_coverage}).}}

\label{tab-stat_each_cat}

\begin{minipage}{0.80\textwidth}
\footnotesize
\begin{tabular}{lcc|ccc|ccc}
\hline
\hline

 &	   &		    &\multicolumn{3}{c}{LR method}			      &\multicolumn{2}{c}{cross-correlation} \\
\hline
 & (1)     & (2)            & (3)                & (4)		& (5)	      & (6)		   & (7)	     \\
 & Catalog &  Radio sources & \% counterparts (\#) & 90\% inside $^a$  & \% spurious        & \% counterparts (\#) & \% spurious     \\
 &         &                &                    & (arcsec)     &	      &       	           &		     \\
\hline
\noalign{\smallskip}
 \multirow{4}{*}{Optical} & U-VIMOS     & 540 & 64\% (347) & 0.7 & 5\% & 64\% (346)   & 12\% \\
                          & v-GEMS      & 646 & 67\% (432) & 0.7 & 5\% & 67\% (435)   & 11\% \\
                          & R-WFI       & 877 & 68\% (600) & 0.7 & 3\% & 67\% (587)   & 6\% \\
                          & z-GOODS     & 164 & 65\% (107) & 0.5 & 4\% & 58\% (95)   & 6\% \\
\noalign{\smallskip}
\hline
\noalign{\smallskip}
 \multirow{4}{*}{NIR    } & H-GNS       &  34 & 70\% (24) & 0.5 & $<$1\% & 62\% (21)   & 6\% \\
                          & H-SOFI      & 523 & 69\% (363) & 0.7 & 2\% & 65\% (339)   & 1\% \\
                          & Ks-ISAAC    & 135 & 81\% (109) & 0.7 & 4\%  & 76\% (102)  & 3\% \\
                          & Ks-MUSYC    & 724 & 77\% (556) & 0.7 & 3\% & 71\% (515)   & 1\% \\
\noalign{\smallskip}
\hline
\noalign{\smallskip}
\multirow{2}{*}{MIR     } & IRAC-SIMPLE & 858 & 87\% (746) & 0.7 & 3\% & 79\% (674)   & 3\% \\
                          & 24um-FIDEL  & 878 & 85\% (745) & 1.1 & 4\% & 79\% (692)   & 2\% \\

\noalign{\smallskip}
\hline

\noalign{\smallskip}
\end{tabular}
$^a$ Radius within 90\% of the counterparts are included.
\end{minipage}
 \end{table*}

\begin{figure}
	\includegraphics[width=\columnwidth]{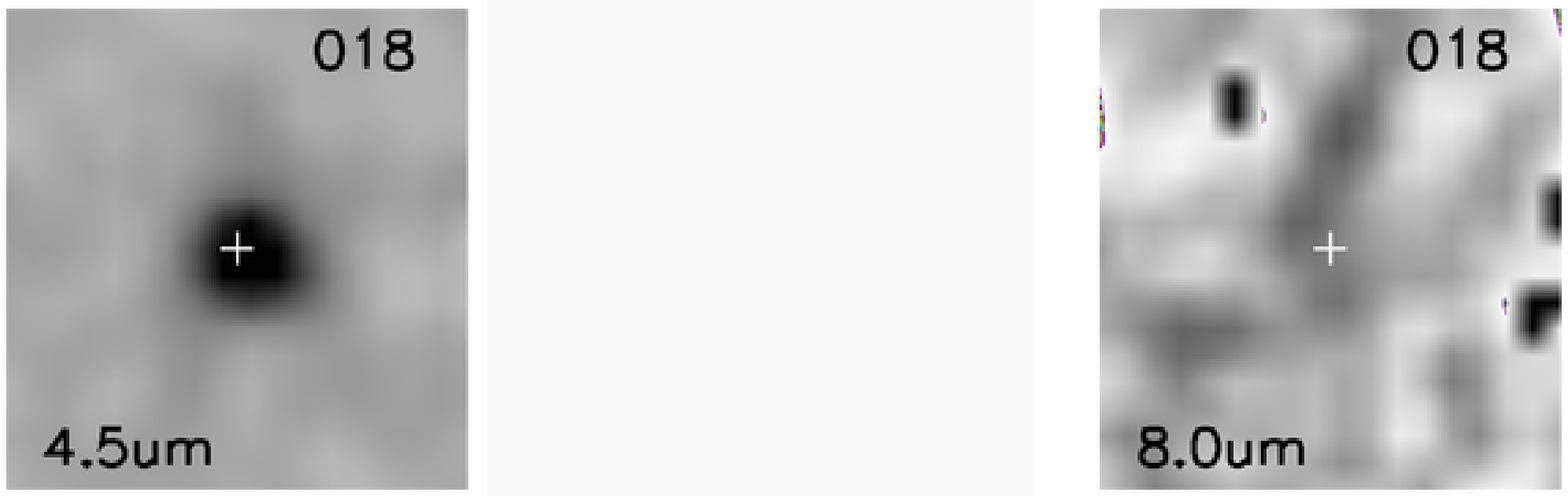} \\  		\includegraphics[width=\columnwidth]{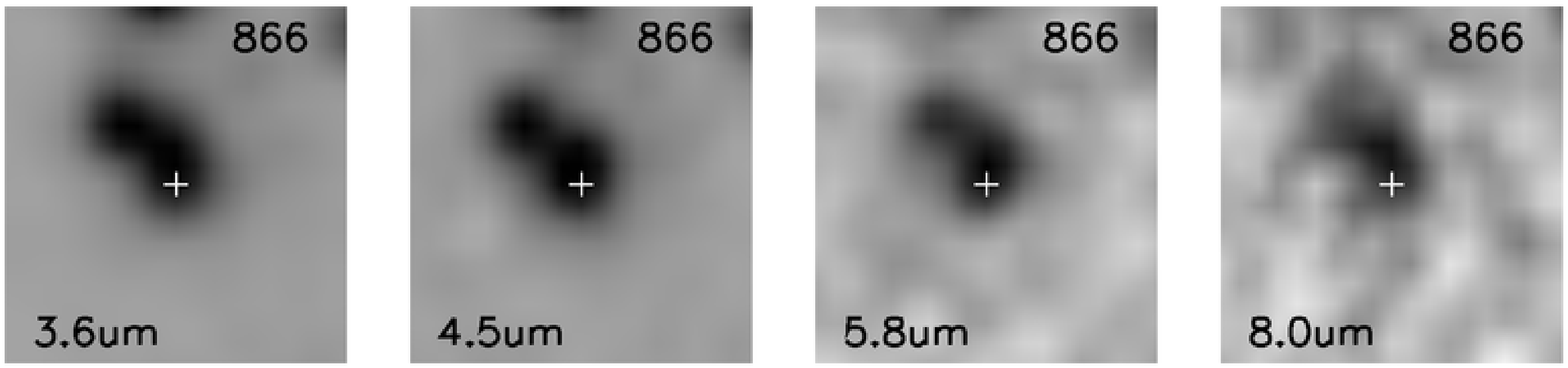} \\
\caption{\small{\textit{Spitzer}/IRAC images of objects RID 018 (top) and 866 (bottom). They represent examples of IRAC sources not present in the SIMPLE catalog whose position has been manually extracted (see Section \ref{sec-revisited} for details): RID 018 has been observed only in two of the four channels, and RID 866 is blended by a nearby source. The position of the radio source is marked by a cross. Each image is 10 arcsec on a side.}}
\label{from_IRAC_img}
\end{figure}

\begin{table*}
      \caption{\small{Radio information and identification process results. A description of the table content is given in Section \ref{sec-final-cat}}. (This Table is available in its entirety in a machine-readable form in the online journal. A portion in shown here for guidance regarding its form and content.)}
\label{tab_final}
\begin{minipage}{0.99\textwidth}
\footnotesize
\begin{tabular}{r c c r@{$\pm$}l c c c c c c }
\hline
\hline
 (1) & (2) & (3) & \multicolumn{2}{c}{(4)} & (5) & (6) & (7) & (8) & (9) & (10) \\
 RID & RA radio & Dec radio & \multicolumn{2}{c}{Sr} & S/N & RA counterpart & Dec counterpart & Reliability & Distance &  Counterpart  \\
 & (J2000) & (J2000) & \multicolumn{2}{c}{($\mu$Jy)} & & (J2000) & (J200) &  & (arcsec) & catalog\\
\hline
\noalign{\smallskip}
360 & 3:32:13.09 &  -27:43:50.9 & 1380.0& 26.6 & 115.8
 & --- & --- & --- & --- & unidentified\\
361 & 3:32:13.24 &  -27:40:43.7 & 58.9& 6.4 & 9.0
 & 03:32:13.24 & -27:40:43.39 & 0.97 & 0.2 & R-WFI\\
362 & 3:32:13.25 &  -27:42:41.3 & 86.4& 6.4 & 13.5
 & 03:32:13.25 & -27:42:40.86 & 1.00 & 0.2 & z-GOODS\\
363 & 3:32:13.36 &  -27:39:35.2 & 48.5& 6.9 & 7.0
 & 03:32:13.32 & -27:39:35.03 & 1.00 & 0.2 & 24um-FIDEL\\
364 & 3:32:13.41 &  -27:33:04.9 & 48.4& 7.8 & 6.2
 & 03:32:13.39 & -27:33:04.93 & 0.97 & 0.3 & R-WFI\\
365 & 3:32:13.50 &  -27:49:53.1 & 44.0& 6.4 & 6.9
 & 03:32:13.48 & -27:49:52.82 & 1.00 & 0.1 & z-GOODS\\
366 & 3:32:13.61 &  -27:34:04.3 & 102.6& 18.5 & 8.3
 & 03:32:13.58 & -27:34:04.37 & 1.00 & 0.4 & R-WFI\\
367 & 3:32:13.65 &  -28:01:01.2 & 35.7& 6.9 & 5.1
 & 03:32:13.62 & -28:01:01.06 & 1.00 & 0.3 & v-GEMS\\
368 & 3:32:13.85 &  -27:56:00.3 & 44.5& 6.4 & 6.9
 & 03:32:13.85 & -27:55:59.95 & 1.00 & 0.2 & 24um-FIDEL\\
369 & 3:32:14.17 &  -27:49:10.6 & 95.4& 6.4 & 14.9
 & 03:32:14.14 & -27:49:10.09 & 0.98 & 0.4 & U-VIMOS\\
370 & 3:32:14.46 &  -27:45:40.8 & 34.9& 6.2 & 5.6
 & 03:32:14.43 & -27:45:40.72 & 1.00 & 0.3 & z-GOODS\\
371 & 3:32:14.60 &  -27:43:05.8 & 32.1& 6.4 & 5.0
 & 03:32:14.60 & -27:43:06.10 & 9.00 & 0.3 & manual\\
372 & 3:32:14.69 &  -28:02:20.2 & 44.9& 7.1 & 6.0
 & 03:32:14.65 & -28:02:19.97 & 1.00 & 0.4 & v-GEMS\\
373 & 3:32:14.85 &  -27:56:40.9 & 109.7& 6.5 & 16.9
 & 03:32:14.83 & -27:56:40.49 & 1.00 & 0.2 & v-GEMS\\
374 & 3:32:15.17 &  -28:05:22.7 & 50.8& 7.9 & 6.3
 & 03:32:15.14 & -28:05:22.24 & 1.00 & 0.3 & R-WFI\\
375 & 3:32:15.34 &  -27:50:37.6 & 43.1& 6.4 & 6.7
 & 03:32:15.32 & -27:50:37.25 & 1.00 & 0.1 & H-GNS\\
\noalign{\smallskip}
\hline
\noalign{\smallskip}
\end{tabular}
\end{minipage}
\end{table*}

In Table \ref{tab_final}, we report the complete list of the
counterparts of the radio sources (see Section
\ref{sec-final-cat}). The counterpart position is taken from an
optical catalog, when available, since these observations have the
highest spatial resolution. In particular, we chose the catalog in
which the counterpart has the highest reliability. According to this
criterion, we selected 104 counterparts from U-VIMOS, 150 from v-GEMS,
301 from R-WFI, and 96 from z-GOODS.  If there was no optical
counterpart above the reliability threshold, we used the coordinates
of the most reliable counterpart found in the NIR catalogs. This happened for 
4 sources from
H-GNS, 24 from H-SOFI, 4 from Ks-ISAAC, and 47 from Ks-MUSYC.  For the
remaining counterparts, we used the position from the IRAC-SIMPLE
catalog (25 sources) and from the 24um-FIDEL one (74 sources).
Finally, there are 10 radio sources (RID: 18, 19, 36, 371, 430, 457,
463, 698, 795, and 866) whose counterpart is clearly visible in the IRAC
images but is not listed in the SIMPLE catalog (or in any other
catalog). The reason is that, since the SIMPLE catalog is extracted  from the combined 3.6 and 4.5 $\mu$m images, either the source was observed only in the first or the second IRAC channel and therefore  not included in the catalog, or it was not
deblended from a nearby object (see Fig. \ref{from_IRAC_img}). In
these cases we have extracted the position of the counterpart from the
IRAC image.

\begin{figure*}
\includegraphics[width=18cm]{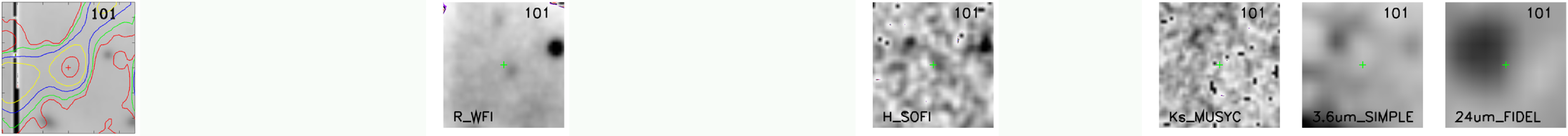}\\
\includegraphics[width=18cm]{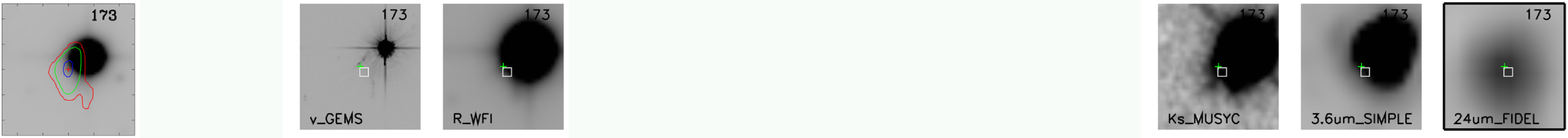}\\
\includegraphics[width=18cm]{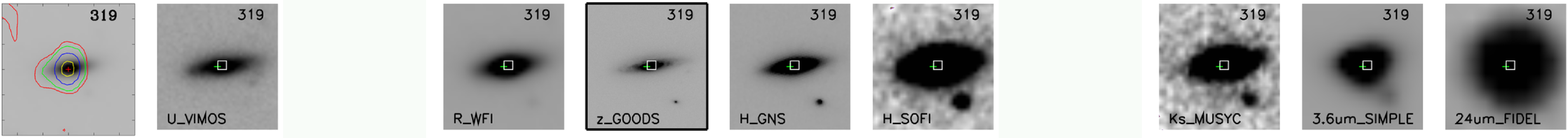}\\
\includegraphics[width=18cm]{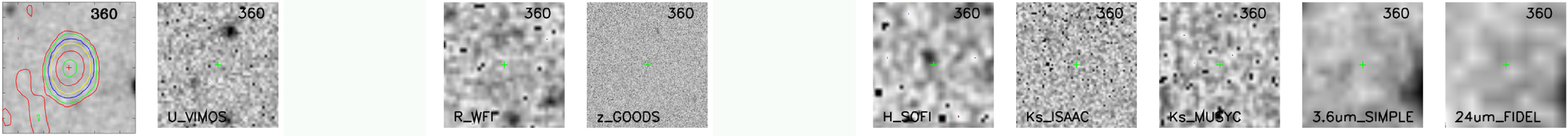}\\
\includegraphics[width=18cm]{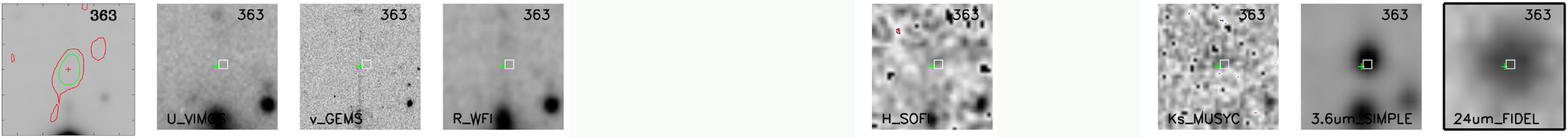}\\
\includegraphics[width=18cm]{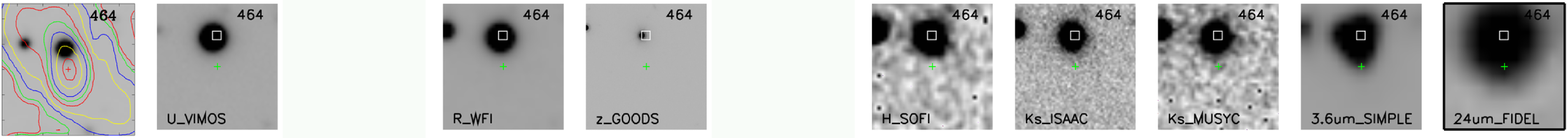}\\
\includegraphics[width=18cm]{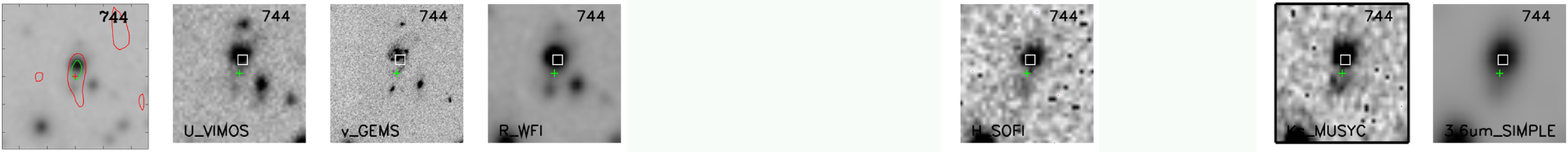}\\
\caption{\small{\textit{Left panels:} radio contours plotted over a 20$\times$20 arcsec R band image. The RID is shown on the top right of each image. \textit{Other panels:} cutouts of the images in the various bands centered at the position of the radio source (marked by a cross).  Each image is 10 arcsec on a side. The corresponding catalog is indicated on the bottom left. The square indicates the position of the selected counterpart. The image from which the counterpart is selected has a black border.}}
\label{fc_examples}
\end{figure*}

As a further check, we extracted 10$\times$10 arcsec cutouts centered
at the radio source position of the images in the various bands, to
visually inspect the counterpart associations. Examples are presented
in Fig. \ref{fc_examples} 
  where the position of
the radio source and of its counterpart are marked by a cross and a
square, respectively.  In the left panels, radio contours are plotted
over a 20$\times$20 arcsec R band image, after the latter has been
registered on the astrometric frame of the radio image. This larger size is chosen for a better view of the radio contours.
In most cases
the selected counterpart is clearly visible in one or more cutouts.
In 12 cases, we found a more convincing counterpart and therefore we
revisited the association; these cases are discussed in more details
in Section \ref{sec-revisited}.

A total of 44 sources are unidentified: most of them are either very faint
radio sources or lie at the edge of the field, where the
multi-wavelength coverage is less rich. They are blank fields in
all the available images (see, e.g., RID 360 in Fig. \ref{fc_examples}). 
We expect only few of them to be spurious radio detections since 
the radio catalog is based on a mosaic image and therefore each object was observed by more than one pointing (N. Miller et al., 2012, in preparation). 
If an object were spurious and due to instrumental effects (e.g., a sidelobe of a nearby bright source) it would not be the same in each pointing. 
Similarly, if it were just a chance noise spike you would expect to see it in only one pointing. 
Therefore we perform source fits
in the individual pointings for the unidentified radio sources with low S/N and we believe that the radio detections are real, with one or two possible exceptions.

In summary, we found a reliable counterpart for 839 out of 883 radio
sources (95\%).

\footnotesize
\normalsize

\subsection{Multiple component radio systems}
\label{sec-multiple}
\begin{figure*}
\centering
\includegraphics[trim=5cm 9cm 5cm 2cm, clip=true, width=12cm]{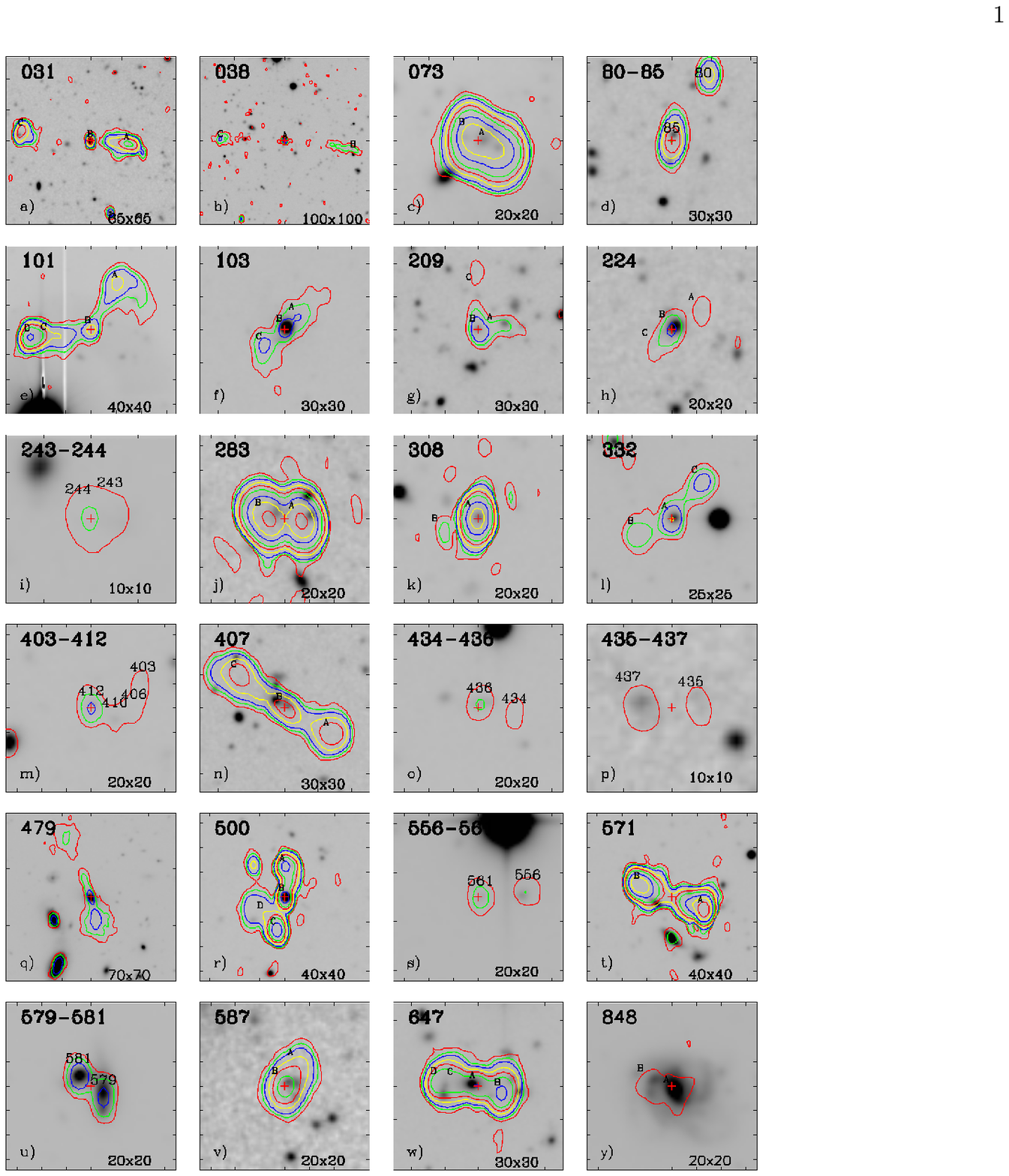}
\caption{\small{Cutouts of radio sources with complex radio morphology. The majority of them ($17/24$) are confirmed as multiple-component system, while for the remaining we find reliable counterparts for each radio component (see Section \ref{sec-multiple} for details). Radio contours are plotted over the R band WFI image. The scale of the cutout is given in arcsec on the bottom right and the RID on the top left.}}
\label{contours_multiple}
\end{figure*}

A multi-wavelength approach is crucial to identifying multiple-component systems. Indeed, the analysis of the radio morphology alone cannot distinguish between pairs of radio sources which are close in projection, or physically connected radio components of the same source.
In our sample there are 24 systems, whose radio morphology can be interpreted as multi-component radio sources. Their radio contours are plotted over the R image in Fig. \ref{contours_multiple}.

We perform the likelihood ratio analysis described in Section \ref{sec-ident} to look for a possible counterpart of each single radio component.
For seven such systems (panels (d), (i), (n), (o), (p), (s), (u) in Fig. \ref{contours_multiple}), we find highly reliable counterparts for each component. Therefore, we claim that they represent distinct sources.
The other 17 are confirmed to be multiple-component systems. 
They have extended radio emission, in most of the cases characterized by a core (not always visible in the radio) and radio
lobes. 
The radio lobes have usually comparable radio power and are not associated with any optical or IR counterpart.
There are some cases where we cannot exclude a possible
contribution to the radio flux density from a superimposed unrelated object
like in sources with RID 38, 73, 209, 283, and 647. 
The complete list of these system with the properties of each radio component is given in N. Miller et al. (2012, in preparation). 
In Appendix \ref{sec-particular} we discuss some peculiar sources.

\subsection{Revisited associations}
\label{sec-revisited}
In section \ref{sec-ident}, we described our method to select the optical and IR counterpart for the radio sources using a likelihood ratio technique. Visually inspecting the results of the identification process, we confirm the association obtained following this procedure in 99\% of the cases. 
In this section, we describe the reasons why we revisited the counterpart association for some peculiar sources. 
In these cases, the most likely counterpart has a reliability under our threshold and was therefore rejected. The two main reasons for this are: (i) the radio source is in a crowded field and therefore all the possible counterparts have low reliability (RID 70, 407, 417, 458, 561, and 797). We based our choice of the counterpart on the radio morphology and on the overall object properties in the various bands (as an example, see the notes on RID 407 in Appendix \ref{sec-particular}). (ii) The radio source is extended, hence the exact position of the radio emission is not well determined or does not correspond to the optical/IR peak emission (RID 407, 420, 521, 804, 828, and 830).
As a consequence, there is an offset between the position of the radio source and the counterpart that has therefore a low reliability.

\subsection{Estimation of spurious associations and comparison with the cross-correlation method}
\label{sec-surious}
For each auxiliary catalog, we estimate the rate of spurious
associations by randomly shifting the position of the radio
sources and computing again the reliability for all the possible
counterparts. We apply only shifts between 5 and 15 arcsec in order
not to exceed the field coverage. We then compute the likelihood ratio value for each one of the shifted sources using equation \ref{equation_LR}, where $q(m)$ and $n(m)$ are the probability distributions derived for the original catalog (see Section \ref{sec-ident}).  
The same reliability threshold of 0.6 is adopted. The
average fraction of false association over 50 different shifts is
reported in Table \ref{tab-stat_each_cat} for each auxiliary catalog. We find spurious fractions
from 3\% up to 5\% for the deep optical catalogs. In the case of H-GNS
the fraction is very low but it could be underestimated due to the
small area covered by this catalog and the consequent small
statistics.

We compare these results with the number of spurious
associations obtained using a simple cross-correlation method. The matching radius chosen for this test is equal to
the radius which includes 90\% of the counterparts
identified with the likelihood-ratio method. These radii are listed in
column (4) of Table \ref{tab-stat_each_cat}. We find that the two methods identify a similar fraction of
sources for the optical catalogs and a somewhat lower one for the IR catalogs. We
estimate the fraction of spurious associations similar to what has
been done for the likelihood method, namely shifting the radio catalog
with respect to the auxiliary one. We used the same set of
displacements as in the previous case. As shown in Table
\ref{tab-stat_each_cat}, our likelihood ratio technique is generally less
affected by spurious contamination especially when applied to the deep
optical catalogs. In particular, in the cases of the U-VIMOS and
v-GEMS catalogs the spurious fraction exceeds 10\% with the
cross-correlation method. If we decrease the searching radius from 0.7 to 0.5 arcsec for all the optical catalogs, the fraction of
false counterparts becomes lower (8\%, 7\%, 3\%, for the U-VIMOS, v-GEMS
and R-WFI catalogs, respectively), but we also miss a significant
fraction of real identification (19\%, 23\% and 14\%, respectively). 
For NIR and MIR catalogs, the two methods are almost equivalent. We
obtain slightly higher fractions of fake associations with the
likelihood ratio technique but with the cross-correlation method we
miss a larger number of counterparts. This is mainly due to the lower source surface density with respect to the optical catalogs.
We note that the shift-and-rematch method tends to overestimate the number of false matches as it ignores the fact that there are a large fraction of the sources that do have counterparts \citep[see][for details.]{broos07,broos11,xue11}. Our estimates should therefore be considered as upper limits. Since this effect is the same both for the likelihood and the cross-correlation method, it does not affect our conclusions.
Finally, we assume that the fraction of spurious association in
the final catalog is equal to the weighted average of the spurious
fraction of each catalog, using the number of counterparts selected
from each catalog as weight. We then expect at most 4\%
spurious counterparts.

\subsection{Comparison with previous work}
\label{sec-comparisonM08}
\begin{figure}
	\centering
	\includegraphics[width=\columnwidth]{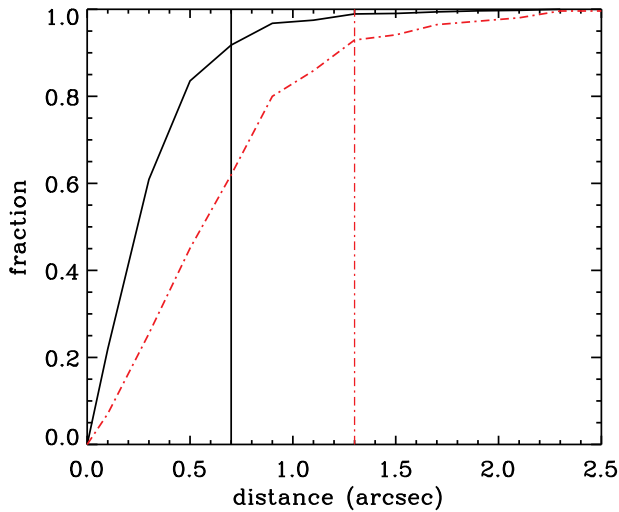}
 \caption{\small{Cumulative distribution of the separation between a radio source and its counterpart for the radio catalog presented in this work (solid line) and in \citet{mainieri08} (dot-dashed line). The vertical lines mark the separation within 90\% of the counterparts are located.}}
 \label{dist_comul}
\end{figure}

The brighter sources of the present catalog were already included in the radio catalog described in \citet{kellermann08}.
We have compared the counterparts found in \citet{mainieri08} (M08,
hereafter) with those selected in this work for the sources in common. We find that in 90\% of the cases the same counterpart is selected. For the remaining 10\%, we select
a different counterpart compared to M08. In most cases the
counterparts were identified in the optical band by M08 while we find a
more convincing counterpart in newly acquired MIR observations. For
eight out of twelve objects that were unidentified or under the threshold
in M08, we now have a reliable counterpart. One example is RID 625 (ID
in the \citet{kellermann08} catalog KID=202) that was in an empty field
in M08 while we identify it at 24 $\mu$m. 
The criteria to identify
the best counterpart candidate of a radio source are slightly different
between the two works. In M08 a likelihood ratio threshold of 0.2 was
adopted, while we use a cut in reliability to take into account the
presence of multiple counterpart candidates of the same radio
source. Our reliability threshold of 0.6 is a bit more conservative
(it correspond to a $LR$ threshold of $\sim 0.3$) and it is aimed at
reducing the number of spurious identifications.  Another difference
between the two works is that in M08 the best counterpart was chosen
according to an a priori ranking of the auxiliary catalogs. The
priority was set according to the depth of the optical/NIR survey and
to the wavelength. 
Given the larger number of
auxiliary catalogs used in this work we use a different
approach: between catalogs in the same wavelength range (optical, NIR
and MIR), we select the counterpart with the highest reliability (see
Section \ref{sec-ident}). This allows us to fully exploit the
information given by the probability distributions obtained with the
likelihood ratio technique. Moreover, we minimize the number of
tentative associations selected in a high priority optical/NIR catalog
yet with a reliability just above the threshold. As a consequence, with our new
approach 90\% of the counterparts have a reliability greater than
0.96, in contrast to 0.83 obtained in M08
(see Fig. 4 in M08).
Moreover, we observe a significant decrease in the average
separation between the radio source and its counterpart compared to M08.
In this work, we find 90\% of the counterparts within 0.7 arcsec
around the radio position, which is about half of the radius found in
M08 (see Fig. \ref{dist_comul}). This is partially due to the change in resolution of the two radio surveys (from 3.5$\arcsec \ \times 3.5\arcsec$ in \citet{kellermann08} to 2.8$\arcsec \ \times 1.6 \arcsec$ in this work). 

In summary, using the new MIR imaging in the E-CDFS area, we
reach the same fraction of identification (95\%) as in M08, although
we adopt a higher reliability threshold and a larger fraction of the
radio sources is in the outer part of the field, where the 
multi-wavelength coverage is poorer. There is only one source
previously identified that we now consider unidentified: RID 101
(see details in Appendix \ref{sec-particular}).

\begin{table*}
 \centering
\footnotesize
 \caption{List of the spectroscopic catalogs considered in this work. The instrument used to obtain the spectra is reported in column (2) and details on the observation and data reduction can be found in the references given in the first column. The label in column (3) are used in the final catalog to identify the source for the spectroscopic redshift. The last column reports the number of spectroscopic redshift adopted from each catalog in this work.}
\label{tab_z-spec}
\begin{tabular}{lccc}
\hline
\hline

 (1)     & (2) & (3) & (4)\\
 Reference & Instrument & Label & Number of z-spec adopted \\
\hline
\noalign{\smallskip}
 This work & VIMOS & P81 & 13\\
 \cite{szokoly04} & FORS1/FORS2 & S04 & 38\\
 \cite{vanzella08} & FORS2 & FORS & 20 \\
 \cite{silverman10} & VIMOS & VJB & 37\\
 \cite{silverman10} & VIMOS & P80 & 3\\
 \cite{silverman10} & Keck & K07 & 32\\
 \cite{silverman10} & Keck & K08 & 18\\
 \cite{treister09} & VIMOS & T09 & 20\\
 \cite{balestra10} & VIMOS-LR & VLR & 24\\
 \cite{balestra10} & VIMOS-MR & VMR & 23\\
 \cite{lefevre04} & VIMOS & VVDS & 19\\
 \cite{ravikumar07} & VIMOS & R07 & 12\\
 \cite{szokoly04} & FORS1/FORS2 & S04F & 1\\
 J. Kurk et al., 2012 (Submitted) & FORS2 & GMASS & 4\\
 \cite{norris06} & 2dF & N06 & 10\\
\noalign{\smallskip}
\hline
\noalign{\smallskip}

\end{tabular}
\end{table*}

\section{Redshift associations}
\label{sec-redshift}

\subsection{New VIMOS spectra and redshifts}
\label{vimos_spectra}
We acquired new optical spectra with the Visible Multi-Object
Spectrograph \citep[VIMOS;][]{lefevre03} at Very Large Telescope (VLT). We carried out one
pointing in the central region of the VLA survey using the
low-resolution (LR) blue grism (R=180, dispersion=5.7 \AA
pixel$^{-1}$) that covers the wavelength range 3700-6700 \AA. The
total exposure time of five hours was set to identify faint optical
counterparts to a limiting point-source magnitude of R$\approx
25$. The mask was designed with the VIMOS Mask Preparation Software
\cite[VMMPS;][]{bottini05} that optimizes the slit assignments based
on our input catalog. We observed a total of 32 VLA sources. The data were reduced using the VIMOS
Interactive Pipeline and Graphical Interface
\citep[VIPGI;][]{scodeggio05}, and the redshifts were estimated using
the EZ\footnote{http://cosmos.iasf-milano.inaf.it/pandora/EZ.html}
software that cross-correlates each spectrum with a template spectrum,
and via visual inspection to validate the result. We derived a
spectroscopic redshift for 13 VLA sources for which previously we had
only a photometric redshift estimate.
The spectra of these
13 radio sources are shown in Appendix \ref{spectra_images}.

\subsection{Spectroscopic redshifts}
\begin{figure}
	\centering
	\includegraphics[width=\columnwidth]{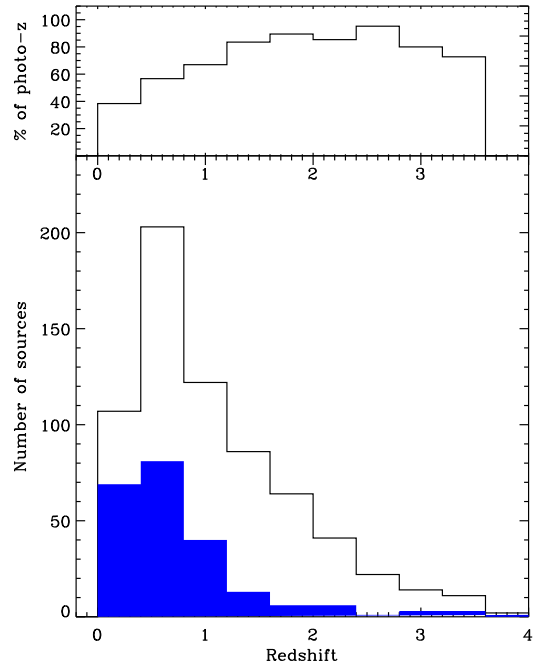}
 \caption{\small{\textit{Top panel:} fraction of photometric redshifts as a function of redshift. \textit{Bottom panel:}  total (empty) and spectroscopic (filled) redshift distributions. The plot is cut at $z=4$ for a better view of the range where both photo-z and spec-z are available.}}
 \label{pz_sz_distr}
\end{figure}
\begin{figure}
	\centering
	\includegraphics[width=\columnwidth]{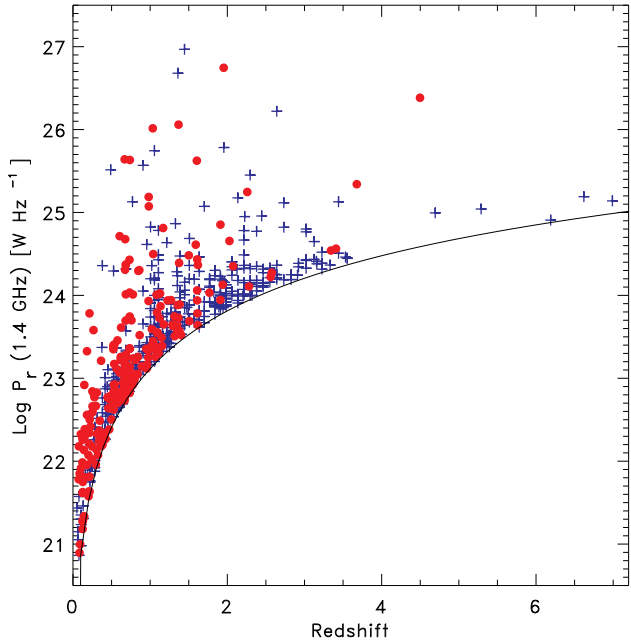}
 \caption{\small{Radio power as a function of redshift. Circles indicate spectroscopic redshifts while crosses indicate photometric redshifts. The solid line shows the radio flux density limit of the VLA survey.}}
 \label{Pr_vs_z}
\end{figure}
\begin{figure}
	\centering
	\includegraphics[width=\columnwidth]{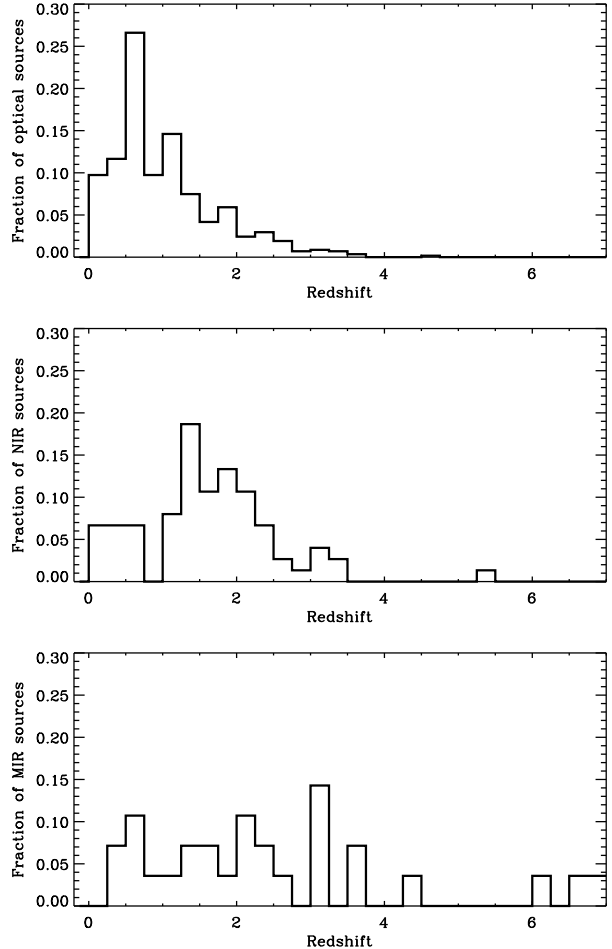}
 \caption{\small{Normalized redshift distribution for the radio sources with an optical counterpart (top), with a NIR counterpart (middle) and with a MIR counterpart (bottom). The mean redshift is increasing from 1 for optical identified sources to 2.5 for the MIR ones (see Section \ref{sec-z-distr}). }}
 \label{z_bands_distr}
\end{figure}
Many spectroscopic campaigns have been conducted in the E-CDFS. A
complete reference list for those used in this work can be found in
Table \ref{tab_z-spec}.  We combine the publicly available
redshifts with our own newly acquired spectra (see
Sec. \ref{vimos_spectra}). We assign a
quality flag (QF) to each redshift by mapping the ones in the original catalogs to a
uniform scale. We use QF=3 to indicate a secure redshift, QF=2 for
reasonable redshift, and QF=1 for tentative redshift or for single
line detection.  The coordinates of the counterparts, identified as
described in Sec. \ref{sec-ident}, are cross-correlated with the
sources reported in the spectroscopic catalogs within 0.2 arcsec. Such a small radius is chosen to minimize the confusion with nearby
sources.  We find a spectroscopic redshift for 274 sources. If a
source has a match in more than one spectroscopic catalog, we verify
the consistency between the corresponding redshifts. For 22 sources,
multiple spectroscopic redshifts differ by more than 0.1. In all but
three of these cases, the QF of the spectroscopic measurements allows
us to select the highest quality $z$. For sources RID= 83, 569,
and 706, where the QF in the various spectroscopic catalogs are
equivalent, we visually checked the spectra to select the more
reliable redshift value. The spectroscopic redshift associated with each
radio source is reported in the column 7 of Table
\ref{tab_counterparts}, with the QF, and the reference in columns 8
and 9, respectively. In summary, 33\% of the radio sources with
counterpart have a spectroscopic redshift, 74\% of which are secure
redshift (QF=3), 18\% have QF=2 and 8\% have a tentative redshift
measurement (QF=1).

\subsection{Photometric redshifts}
In order to increase the redshift completeness of our sample, we
also use photometric redshift estimates. We use the photometric
redshift catalog compiled by \citet{luo10} and \citet{rafferty11}.  These redshifts are based on a large number of
photometric bands: the COMBO-17 optical catalog
\citep{wolf04,wolf08}, the GOODS-S MUSIC catalog \citep{grazian06},
the MUSYC BVR-detected catalog \citep{gawiser06}, the deep GOODS-S
VIMOS U-band catalog \citep{nonino09}, the GALEX Data Release
4\footnote{http://galex.stsci.edu/GR4/}, the MUSYC near-infrared
catalogue \citep{taylor09}, and the SIMPLE mid-infrared one
\citep{damen11}.  Starting from this photometric data set, the
publicly available Zurich Extragalactic Bayesian Redshift Analyzer  \citep[ZEBRA;][]{feldmann06} code was used to derive photometric
redshifts via a maximum likelihood technique. The set of templates
used includes: 259 galaxy templates constructed from PEGASE stellar
population synthesis models, a set of hybrid (galaxy+AGN) templates
and ten empirical AGN templates from \citet{polletta07}. We refer the
reader to \citet{luo10} and \citet{rafferty11} for a
more detailed description of the procedure adopted to estimate
photometric redshifts. 
We cross-correlate the photometric redshift catalog with the radio
source counterparts, selected as described in \ref{ident_results}.
Given the high background surface density distribution of the
photometric catalog, we adopt a matching radius of 0.2
arcsec. This way we minimize the risk of associating to the radio
counterpart the redshift of a nearby source.  We find 623 matches out
of the 839 identified radio sources. The mean separation between the
radio counterparts and the corresponding photometric redshift
coordinates is 0.03 arcsec.
For the remaining 216 objects we consider three other compilations of
photometric redshifts: the MUSYC-E-CDFS catalog \citep{cardamone10},
the GOODS-MUSIC catalog \citep{santini09}, and the K-selected MUSYC
catalog \citep{taylor09}. The latter is based on a NIR selected
catalog, and therefore is particularly useful to assign a redshift
to radio sources with no counterpart in the optical. We find 13, 4,
and 35 additional redshifts, respectively. In Summary, we associate
a photometric redshift to 673 (80\%) out of the 839 identified radio
sources.

For the sub-sample with available secure spectroscopic redshift (QF=3),
we compute the normalized median absolute scatter,
\begin{equation}
\sigma_{NMAD}= 1.48 \times median\left(\frac{|\Delta z - median(\Delta
  z)|}{1+z_{spec}}\right),
\end{equation}
where $\Delta z = z_{phot}-z_{spec}$, which is an estimate of the
quality of the photometric redshift which is less sensitive to outliers than
the standard deviation \citep{brammer08}. 
We find $\sigma_{NMAD}=0.01$
that is comparable, and even slightly better, to what found for the same indicator in the photometric redshift catalogs considered \citep{santini09,taylor09,cardamone10,rafferty11}.
Therefore, we conclude that the accuracy of the photometric redshifts for our radio selected sample is comparable to what estimated for the overall population in the photometric catalogs.  
We note that this indicator assumes that the spectroscopic sub-sample is representative of the full sample. This assumption is 
likely not entirely true and consequently it gives an overestimation of the accuracy of the photo-z for the whole sample. \citet{luo10} estimated that the uncertainties on the photometric redshifts for the full sample are a factor of six higher.
We also note that the photometric redshift errors given in column 6 of Table \ref{tab_counterparts} from the \citet{rafferty11} catalog are known to be underestimated (see \citet{luo10}). A more realistic estimate is given by the $\sigma_{NMAD}$ parameter, which is around 6\% as discussed above.

\subsection{Redshift distribution}
\label{sec-z-distr}
In the case where both spectroscopic and photometric redshifts are
available, we use the spectroscopic one if QF $\ge$ 2 and the
photometric redshift otherwise. For spectroscopic redshifts with QF=1
we find $\sigma_{NMAD}=0.28$, which we interpret as an indication of
the poor quality of these spectroscopic redshifts. Combining
spectroscopic and photometric information, we assign a redshift to 678
objects, 81\% of the radio sources with counterpart (252 spectroscopic redshifts and 426 photometric redshifts). This fraction
underestimates the redshift completeness of our sample since in the
outermost part of the field there are no redshift measurements
available. Therefore, we restrict our redshift distribution
analysis to the sources in the area covered by the photometric
redshift catalogs\footnote{Only three sources outside this region have
  a spectroscopic redshift.}. This region is plotted with a dot-dashed line in the right panel of Fig. \ref{aux_cat_coverage}.
The number of radio sources included is 779, and 87\% of them have a
redshift.  The total redshift distribution is plotted in
Fig. \ref{pz_sz_distr}, where the filled histogram represents the
distribution of sources with spectroscopic redshifts. The top panel
shows the fraction of photometric
redshifts. We note that photometric redshift measurements become
increasingly important at higher redshifts where optical spectroscopic observations
become more challenging. The mean redshift for the whole sample is
1.1 and the median is 0.9.
Fig. \ref{Pr_vs_z} shows the radio power as a function of redshift for 
sources with either spectroscopic (circle) or photometric (cross) redshift. The solid line represents the flux density limit of the survey. 
If we divide the radio sources based on their identification band, we
observe an increase in the mean (median) redshift from 1.0 (0.8)
for the optical identified sources to 2.5 (2.1) for the MIR ones
(see Table \ref{tab_z_dist} and Fig. \ref{z_bands_distr}).
The statistical significance of the different redshift distributions is examined with the Kolmogorov-Smirnov (K-S) test. The difference between the optical and NIR distributions and that between the optical and MIR distributions are confirmed with a significance level $\gg 99\%$. For the NIR and MIR redshift distributions the K-S test gives a significance of 99\%.
Both
spectroscopic and photometric redshift estimates become more
challenging moving to high redshift objects and this is the reason why
the fraction of sources with a redshift estimate drops from 88\% for
the optical identified sources to 25\% for the MIR ones (see Table
\ref{tab_z_dist}).

\begin{table*}
\footnotesize
       \caption{Main characteristics and redshift information of the radio source counterparts. A description of the table content is given in Section \ref{sec-final-cat}. (This table is available in its entirety in a machine-readable form in the online journal. A portion in shown here for guidance regarding its form and content.)}
\label{tab_counterparts}
\centering
\tiny
\begin{tabular}{r c c c c c c c c c c c}
\hline
\hline
 (1)  & (2) & (3) & (4) & (5) & (6) & (7) & (8) & (9) & (10) & (11) & (13) \\
 RID & R mag & Ks mag & 3.6$\mu$m mag & best-z & photo-z & spec-z & QF & reference & S$_{0.5-2keV}$ & S$_{2-10keV}$ & XID \\
 & (AB) & (AB) & (AB) & & & & & & (erg cm$^{-2}$ s$^{-1}$) & (erg cm$^{-2}$ s$^{-1}$) & \\
 \hline
\noalign{\smallskip}
\noalign{\smallskip}
360 & --- & --- & ---
 & --- & --- & --- & --- & ---
 & --- & --- & --- \\
361 & 25.92 $\pm$ 0.22 & 22.02 $\pm$ 0.18 & 21.02 $\pm$ 0.02
 & 1.39 & $1.39^{+0.02}_{-0.18}$ & --- & --- & ---
 & --- & --- & --- \\
362 & 20.40 $\pm$ 0.00 & 18.47 $\pm$ 0.01 & 18.46 $\pm$ 0.00
 & 0.61 & $0.60^{+0.00}_{-0.00}$ & 0.607 & 3 & VMR
 & (1.58$\pm$0.04)$\times 10^{-15}$ & (1.06$\pm$0.02$)\times 10^{-14}$ & 193\\
363 & --- & --- & 21.33 $\pm$ 0.30
 & --- & --- & --- & --- & ---
 & --- & --- & --- \\
364 & 25.57 $\pm$ 0.14 & --- & 22.50 $\pm$ 0.07
 & 3.02 & $3.02^{+0.06}_{-0.14}$ & --- & --- & ---
 & (1.74$\pm$0.23)$\times 10^{-15}$ & (8.00$\pm$1.05$)\times 10^{-15}$ & 1330\\
365 & 22.78 $\pm$ 0.02 & 19.60 $\pm$ 0.03 & 19.29 $\pm$ 0.00
 & 0.73 & $0.73^{+0.00}_{-0.01}$ & 0.731 & 3 & FORS
 & --- & --- & --- \\
366 & 22.91 $\pm$ 0.02 & 19.98 $\pm$ 0.04 & 19.73 $\pm$ 0.00
 & 0.64 & $0.64^{+0.02}_{-0.02}$ & --- & --- & ---
 & --- & --- & --- \\
367 & 21.17 $\pm$ 0.00 & 19.16 $\pm$ 0.02 & 19.16 $\pm$ 0.00
 & 0.60 & $0.60^{+0.02}_{-0.03}$ & --- & --- & ---
 & --- & --- & --- \\
368 & --- & --- & 21.60 $\pm$ 0.07
 & --- & --- & --- & --- & ---
 & (2.51$\pm$0.77)$\times 10^{-16}$ & (1.59$\pm$0.49$)\times 10^{-15}$ & 197\\
369 & 24.44 $\pm$ 0.06 & --- & 21.40 $\pm$ 0.02
 & 2.08 & $2.06^{+0.02}_{-0.02}$ & 2.076 & 3 & VJB
 & (4.89$\pm$0.71)$\times 10^{-17}$ & (1.51$\pm$0.22$)\times 10^{-16}$ & 202\\
370 & 21.30 $\pm$ 0.01 & 19.14 $\pm$ 0.02 & 19.48 $\pm$ 0.00
 & 0.30 & $0.28^{+0.00}_{-0.00}$ & 0.296 & 3 & VMR
 & --- & --- & --- \\
371 & --- & --- & 24.08 $\pm$ 0.29
 & 6.19 & $6.19^{+0.81}_{-5.03}$ & --- & --- & ---
 & --- & --- & --- \\
372 & 19.87 $\pm$ 0.00 & 18.36 $\pm$ 0.01 & 18.89 $\pm$ 0.00
 & 0.35 & $0.35^{+0.00}_{-0.03}$ & --- & --- & ---
 & --- & --- & --- \\
373 & 20.90 $\pm$ 0.01 & 17.94 $\pm$ 0.01 & 17.62 $\pm$ 0.00
 & 0.73 & $0.73^{+0.00}_{-0.01}$ & 0.733 & 3 & VLR
 & --- & --- & --- \\
374 & 19.04 $\pm$ 0.00 & --- & 17.69 $\pm$ 0.00
 & --- & --- & --- & --- & ---
 & --- & --- & --- \\
375 & --- & --- & ---
 & 1.51 & $1.51^{+0.05}_{-0.07}$ & --- & --- & ---
 & (1.21$\pm$0.13)$\times 10^{-16}$ & (4.43$\pm$0.48$)\times 10^{-16}$ & 217\\
\noalign{\smallskip}
\hline
\noalign{\smallskip}
\end{tabular}
\end{table*}

\begin{table*}
\begin{minipage}{0.99\textwidth}
\footnotesize
       \caption{Redshift distribution of the counterparts divided according to their wavelength selection. The number in column (1) is the number of radio sources whose counterpart is found in the group of catalogs of the corresponding row.}
\label{tab_z_dist}
\centering
\begin{tabular}{lccc|cc}
\hline
\hline
	& (1) 				& (2)	& (3)		&(4) 	& (5)	 \\
    & Identified sources & With z & Fraction  & Mean z & Median z\\
\hline
\noalign{\smallskip}
\noalign{\smallskip}
Optical & 652	& 575 & 88\%  & 1.0 & 0.8 \\ 
NIR & 76	 & 75  & 98\%  & 1.6 & 1.5 \\ 
MIR & 111 & 28  & 25\%  & 2.5 & 2.1 \\ 
\noalign{\smallskip}
\hline
\noalign{\smallskip}
\end{tabular}
\end{minipage}
\end{table*}

\section{X-ray counterparts}
\label{sec-X-ray}
\textit{Chandra} has imaged in the X-ray the area of the E-CDFS as
part of two different programs. The first is a 250 ks exposure
observation that covers almost the whole field (0.28 deg$^2$)
\citep{lehmer05}. The survey reaches sensitivity limits of
$1.1\times10^{-16}$ and $6.7\times10^{-16}$ ergs cm$^{-2}$ s$^{-1}$
for the soft (0.5 -- 2.0 keV) and hard (2 -- 10 keV) bands, respectively.
The second set is a much deeper 4 Ms \textit{Chandra} observation 
covering only the central part of the field ($\approx 0.1$
deg$^2$). The on-axis flux limits are $9.1\times10^{-18}$ for the soft
band and $5.5\times10^{-17}$ ergs cm$^{-2}$ s$^{-1}$ for the hard band
\citep{xue11}. 
We cross-correlated the radio
source catalog with the X-ray ones. Due to the low surface density
of X-ray sources a simple positional match is almost equivalent to the
likelihood ratio technique. The searching radius was set to three times
the sum in quadrature of the errors on the radio and X-ray
positions. In case of multiple counterpart candidates, we selected the one closest to the radio source position.
We find 129 radio sources with X-ray detection from the 4 Ms \textit{Chandra} catalog, and 99 sources from the 250 ks catalog. Combining the two lists, we
have X-ray detection for 25\% of our radio sources. Their flux in the soft and hard band are reported in columns 10 and 11 of Table \ref{tab_counterparts}. We refer the reader to Section
4.2 of \citet{vattakunnel12} for a description of the properties of the
radio sources with X-ray counterpart. In the following section we
focus our attention on the sources for which we obtain only upper
limits on their X-ray flux.
\subsection{Average X-ray properties of radio-only detected  sources}
The majority ($75\%$) of our radio sources has no X-ray
counterpart. Even in the region covered by the 4 Ms \textit{Chandra}
observation, the fraction of identified sources is only $\sim$ 60\%
\citep{vattakunnel12}.
For the radio-only sources we perform aperture photometry on the X-ray images
at the position of the radio source. The X-ray detected sources are
masked and replaced with a Poissonian background based on the value of
the measured local background. The photometry is done separately in
the soft (0.5--2 keV) and in the hard (2--10 keV) bands. To derive the
average properties of these objects, we perform a stacking analysis of
the \textit{Chandra} images. In particular, we stack separately
sources with counterparts selected from an optical catalog, from a NIR
catalog and from a MIR one. The net counts obtained in both hard and
soft X-ray bands are reported in Table \ref{tab-stacking}. The
detection for the optical selected sources is highly significant in
both bands, while for the MIR ones it is only marginal in the soft band. The NIR selected
sources have marginal detection only in the soft band
and are not detected in the hard band. For each group, we evaluate
the average hardness ratio defined as $HR=(H-S)/(H+S)$, where $H$ and
$S$ are the total net counts in the hard and soft band, respectively
(column 4 in Table \ref{tab-stacking}).  In particular we note that
MIR selected sources have a hard hardness ratio, $HR=0.4\pm 0.2$,
supporting the hypothesis that these objects are obscured sources.
This $HR$ value corresponds to an effective X-ray photon indices $\Gamma=0.07^{+0.14}_{-0.13}$.
\begin{table*}
\begin{minipage}{0.99\textwidth}
\footnotesize
      \caption{Net counts and $HR$ for radio sources without X-ray detection. The number in column (1) is the number of radio sources which have a optical/IR counterpart but not detected in the X-ray.}
\label{tab-stacking}
\centering
\begin{tabular}{lcccc}
\hline
\hline
	& (1) 				& (2)			& (3)			&(4)	 \\
    & Number of sources & 0.5--2 keV cts & 2--10 keV cts  & HR \\
\hline
\noalign{\smallskip}
Optical & 426 & 676 $\pm$ 68 & 374 $\pm$ 97 & -0.3 $\pm$ 0.1 \\ 
NIR 		& 59  &  78 $\pm$ 24 & --- & --- \\
MIR 		& 86  &  79 $\pm$ 33 & 227 $\pm$ 50 & 0.4 $\pm$ 0.2 \\
\hline
$\log P_{\rm r} <$ 22.9 		 & 122 & 231 $\pm$ 36 & 162 $\pm$ 51 & -0.2 $\pm$ 0.2\\ 
22.9$\leq \log P_{\rm r} <$23.5 & 121 & 194 $\pm$ 35 & 98 $\pm$ 50 & -0.3 $\pm$ 0.2 \\ 
$\log P_{\rm r} \geq$ 23.5 		& 97 & 120 $\pm$ 31 & 52 $\pm$ 43 & -0.4 $\pm$ 0.4\\ 
\noalign{\smallskip}
\hline
\noalign{\smallskip}
\end{tabular}
\end{minipage}
       
\end{table*}

We also split the sample of X-ray undetected sources in radio power
bins to investigate if there are any changes in the average X-ray
spectral properties as a function of radio power. We consider only
sources with $z < 1.5 $, where we have a more uniform distribution in
radio power, from $10^{20}$ to $10^{27}$ W Hz$^{-1}$. The radio power bins, the net counts, and the
$HR$ are reported in Table \ref{tab-stacking}. We find a roughly
constant value of $HR$ and therefore no significant change in the
average X-ray spectral properties as a function of radio power.

\section{Catalog description}
\label{sec-final-cat}
In this section we describe the catalog containing the results of the
optical and IR counterpart identification process. The information
is divided into two tables.
In Table \ref{tab_final} we include the
radio data from N. Miller et al. (2012, in preparation) that were used in this work and the
results of the identification process. In Table \ref{tab_counterparts}
we list the main characteristics of the optical or IR counterpart,
the redshift information and the X-ray data. The catalog columns are
organized as follows. In Table \ref{tab_final}:

\begin{itemize}
\item (1) Identification number of the radio source (RID).
\item (2) and (3) Right ascension and declination of the radio source.
\item (4) Radio flux density and $1\sigma$ error in $\mu$Jy.
\item (5) Signal to noise ratio.
\item (6)and (7) Right ascension and declination of the counterpart.
\item (8) Reliability of the association\footnote{Sources whose identification has been revisited (see Section \ref{sec-revisited}) or for which the counterpart position has been extracted from the IRAC image (see Section \ref{ident_results}) have $Rel=9$.}.
\item (9) Distance between the radio source and the counterpart in
  arcsec.
\item (10) Catalog from which the counterpart is selected (see Table \ref{tab_aux_cat}).
\end{itemize}

In Table \ref{tab_counterparts}:
\begin{itemize}
\item (1) Identification number of the radio source (RID).
\item (2) R-band AB magnitude of the counterpart from the WFI catalog
  and associated error.
\item (3) K-band AB magnitude of the counterpart from the MUSYC
  catalog and associated error.
\item (4) Flux density at 3.6$\mu$m of the counterpart from the SIMPLE
  catalog and associated error.
\item (5) Best redshift of the counterpart: spectroscopic if $QF\geq 2$, photometric otherwise. 
\item (6) Photometric redshift with upper and lower 68\% confidence
  level.
\item (7) Spectroscopic redshift.
\item (8) Quality flag (QF): 3 for secure redshift, 2 for resonable redshift and 1 for one line detection or tentative redshift.
\item (9) Source of the spectroscopic z (see Table \ref{tab_z-spec}).
\item (10) X-ray soft band flux (0.5–-2.0 keV) and associated error.
\item (11) X-ray hard band flux (2--10 keV) and associated error.
\item (12) X-ray ID (from \citet{xue11} if $<$1000, from \citet{lehmer05} if $>$1000 [ID-Lehmer+1000])
\end{itemize}

\section{Discussion and Conclusions}
\label{sec-discussion}
We have presented the optical and IR counterparts of the radio sources in N. Miller et al. (2012, in preparation) catalog.
The results are collected in a new catalog\footnote{The catalog is available in ASCII format in the on-line material.} containing the counterpart data and the redshift information. 
A detailed characterization of the physical properties of these sources will be presented in M. Bonzini et al. (2012, in preparation). 
This work has demonstrated the difficulties in, and the requirements for, the identification of the sub-mJy radio population. 
The main results of our analyses are as following:
\begin{enumerate}
\item \textit{Importance of multi-wavelength observations.}  We identify the counterparts  for a high fraction (95\%) of radio sources. 
In order to reach such a completeness it is necessary to include not only optical
observations, but also near and far infrared data. Optical surveys alone, even in the deepest fields, allow us to
identify only $\sim70\%$ of the radio sources. With just MIR observations the
fraction rises to 86\%, but it is by only combining the information
from all wavelengths that we reach 95\% completeness.
The multi-wavelength coverage is also important to obtain a high redshift completeness. 
Indeed, only 31\% of our radio sources have spectroscopic information, while the majority have a photometric redshift. 
\item \textit{Importance of the counterpart analysis to confirm multiple-component
  radio sources.}
In this work we have found many examples that show the importance of the combination of radio and optical/IR data to correctly identify multiple-component radio system. In many cases, sources whose radio morphology suggested a complex radio structure (e.g., KID 114) have been identified as independent sources. 
The opposite case is represented by source RID 73. Here, we conclude that the radio emission is associated with a single compact radio-lobe source with a single optical counterpart. 
\item \textit{Comparison between likelihood ratio and
  cross-correlation methods}.  In Sec \ref{sec-surious}, we compare
  the likelihood ratio technique with the positional matching
  method. This work has shown that the latter is hardly applicable to deep optical surveys since it leads to a large fraction of spurious
  matches ($\sim 10\%$).
  With our technique instead the rate of spurious matches is lower due to the exploitation of the information given by the probability distribution of background sources in the optical catalogs. 
  We have also shown that to reach the same level of spurious contamination with the cross-correlation method the fraction of identified sources decreases by $\sim  18\%$.
  At longer wavelengths, i.e. in the NIR and in the MIR, the differences of the two methods are negligible. We find a comparable fraction of expected spurious counterparts and a similar completeness. 
  This is mainly due to the lower background surface density of objects  in the auxiliary catalogs.
  In section \ref{sec-revisited} we point out two cases where the likelihood ratio method can fail in identifying the correct optical counterpart, that is the presence of many close sources around the radio object and extended radio emission. We note that with the cross-correlation method these problems are even more severe; in crowded regions it selects the closest source, regardless of its properties. 
In case of extended radio sources, there could be an offset between the peak of the emission at different wavelengths larger than the searching radius. In this case, the cross-correlation method would not be able to give any counterpart candidate.  
\item \textit{Comparison with M08 work.}
Compared with the sample studied in M08, our sample is about 3 times larger and most of the new sources have a low radio flux density or lie at the edges of the E-CDFS. 
That makes their identification more challenging. 
However, more and deeper catalogs are now available in the E-CDFS and, using these data, we were able to reach the same identification completeness as in M08 for the new sample. 
We find general agreement between the counterparts found in the two works for the radio sources in common.
The main improvement is a more reliable identification in particular of the optically faint radio sources, obtained by adopting a stricter acceptance criteria and giving more importance to the IR selected catalogs (Section \ref{sec-comparisonM08}).
\item \textit{Importance of MIR observations to find the counterpart of
  high redshift or heavily obscured radio objects.}
    Some radio sources (12\%) have a reliable counterpart only in the catalogs based on the \textit{Spitzer} data. These sources are particularly interesting since they are the best candidate high-z objects.
 In Sec. \ref{sec-z-distr}, we describe the redshift distribution of the radio sources divided according to their identification band. 
 Indeed, we find a clear trend for sources identified at longer wavelengths to have higher redshifts, as shown in Fig. \ref{z_bands_distr}.
Moreover, the stacking analysis of the X-ray images of the MIR selected sources, has revealed that they tend, on average, to have hard X-ray spectra ($HR=0.4$).
This supports the idea that they are obscured sources.
\end{enumerate}

\acknowledgments
This work is based on VLA observation. The VLA is a facility of the National Radio Astronomy Observatory operated by associated Universities Inc. under cooperative agreement with the US National Science Foundation. This work is based on observations made with the VIMOS Telescopes at the ESO Paranal Observatories under Program ID(s) 081.A-0525 and 171.A-3045.
 We acknowledge the ESO/GOODS project for the ISAAC and FORS2 data obtained using the Very Large Telescope  at the ESO Paranal Observatory under Program ID(s): LP168.A-0485,
 170.A-0788, 074.A-0709, and 275.A-5060. 
S.V. and P.T. acknowledge support under the contract ASI/INAF I/009/10/0.
W.N.B., B.L., and Y.Q.X. acknowledge the Chandra X-ray Center Grant SP1-12007A  and NASA ADP Grant NNX10AC99G.
Y.Q.X. acknowledges the financial support of the Youth 1000 Plan (QingNianQianRen) program and the USTC startup funding (ZC9850290195).
M.B. acknowledges support from and participation in the International Max-Planck Research School on Astrophysics at the Ludwig-Maximilians University.




\clearpage
\newpage
\appendix

\section{New VIMOS/VLT spectra.}
\label{spectra_images}
Newly acquired VIMOS spectra for the counterparts of 13 VLA sources (see
Sec. \ref{vimos_spectra}). We used the low-resolution blue grism
(R=180), with a total exposure time on source of 5 hours. Each plot of Fig. \ref{fig_spectra}
shows the spectra and the position of the main spectral features. We make bold the names of the lines actually used to identify the redshift. The labels report the source RID, the redshift, and the corresponding quality flag (QF).

\begin{figure*}[ht]
\begin{center}
\includegraphics[width=3.in]{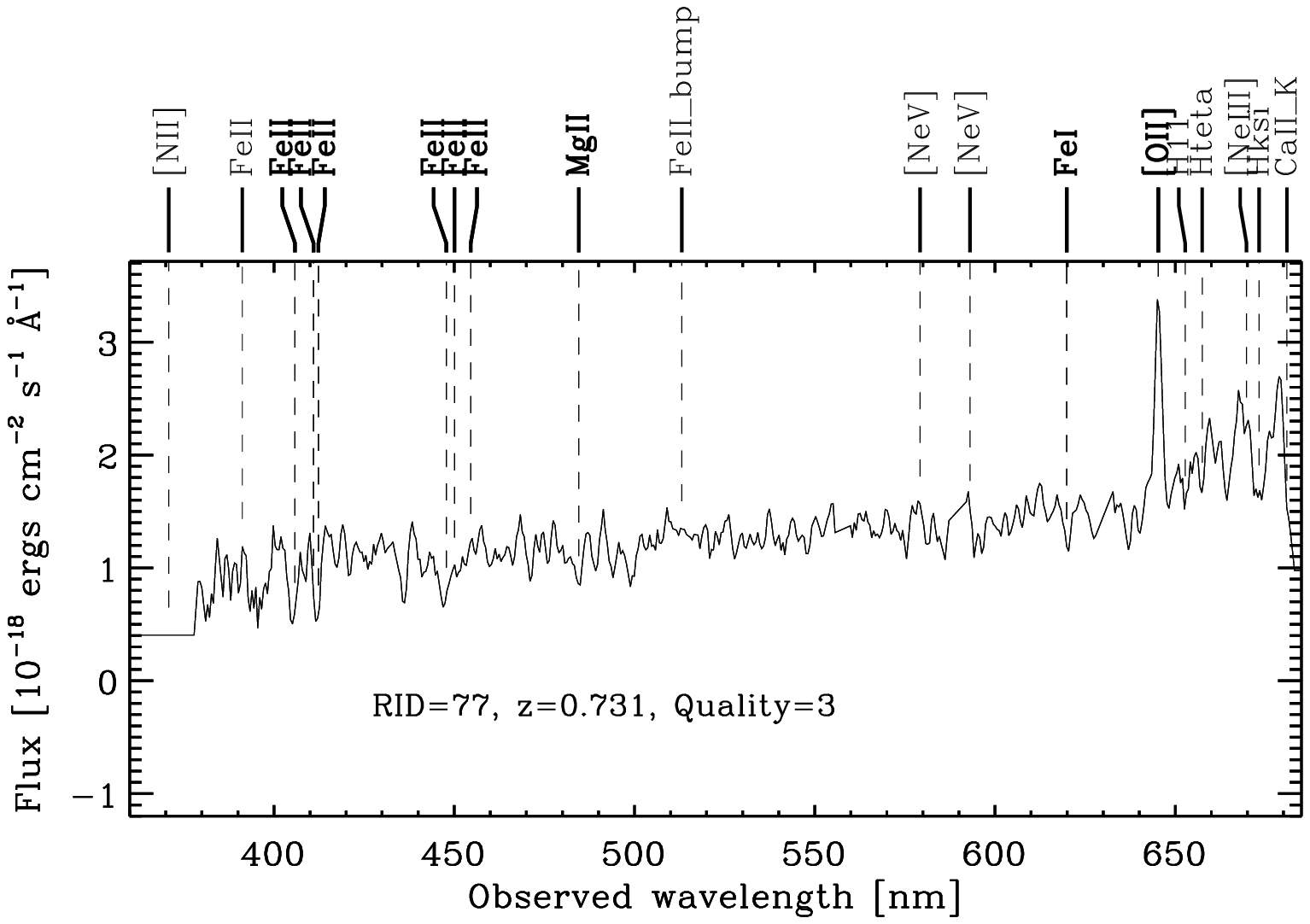}
\includegraphics[width=3.in]{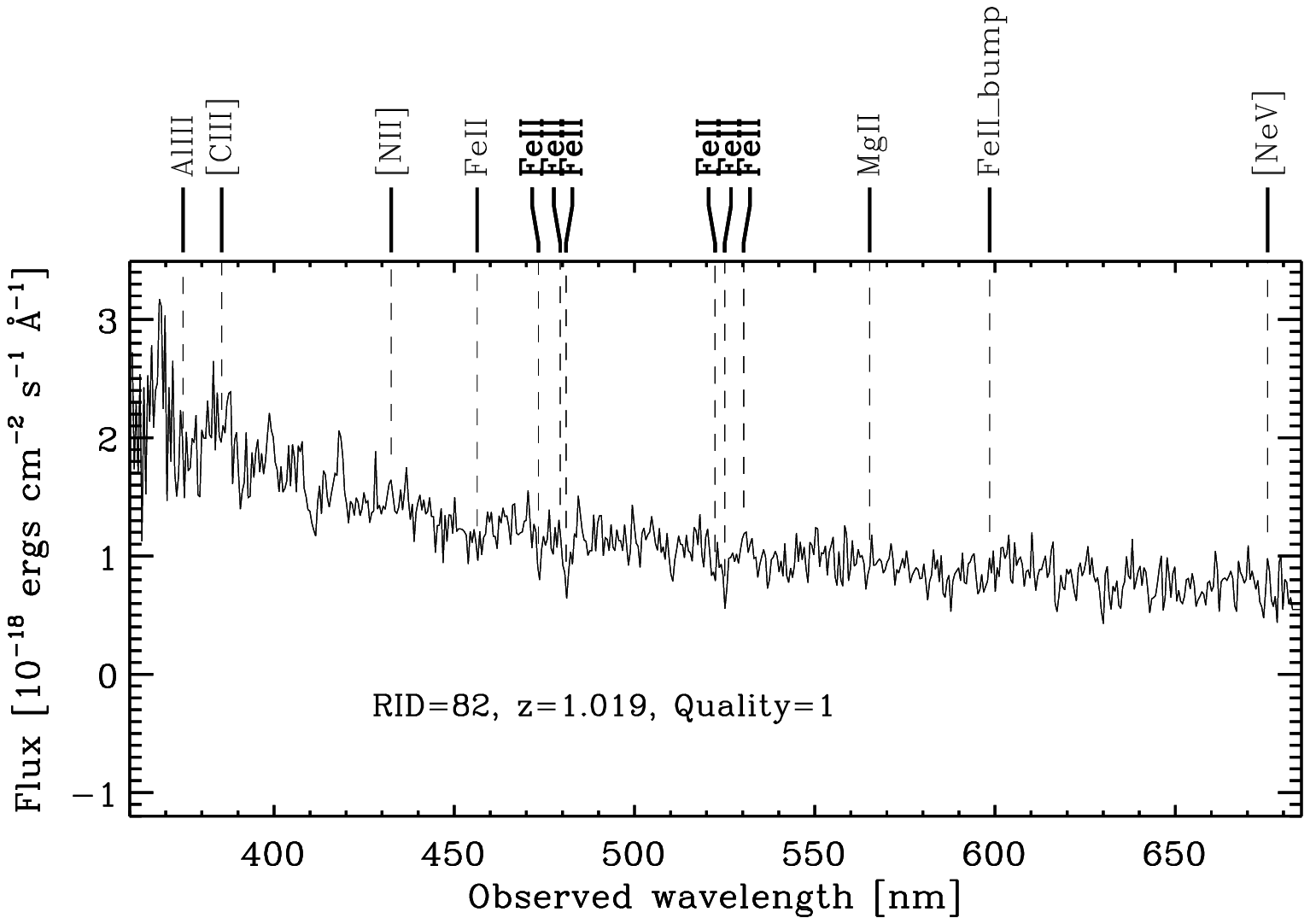}\\
\includegraphics[width=3.in]{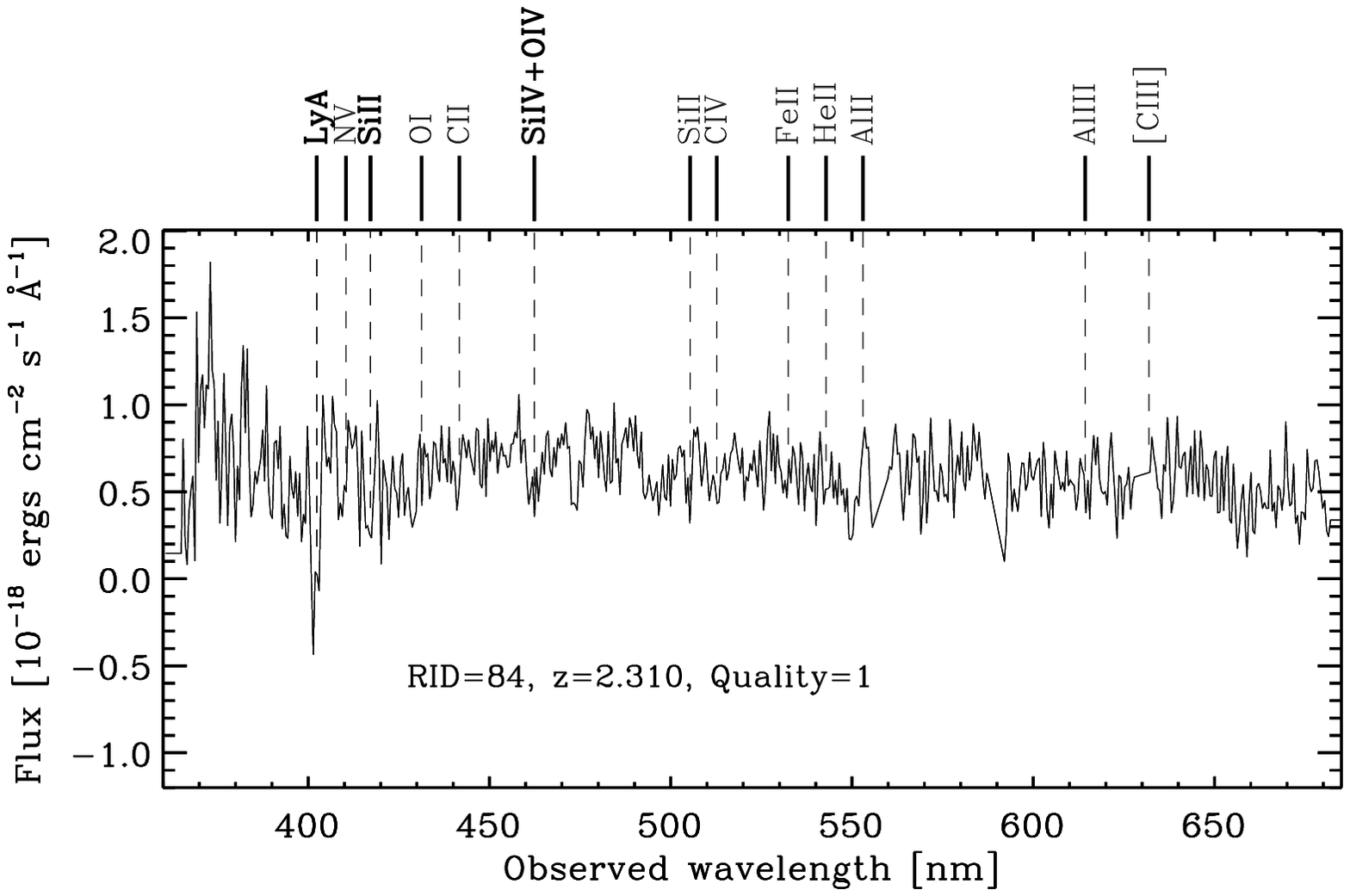}
\includegraphics[width=3.in]{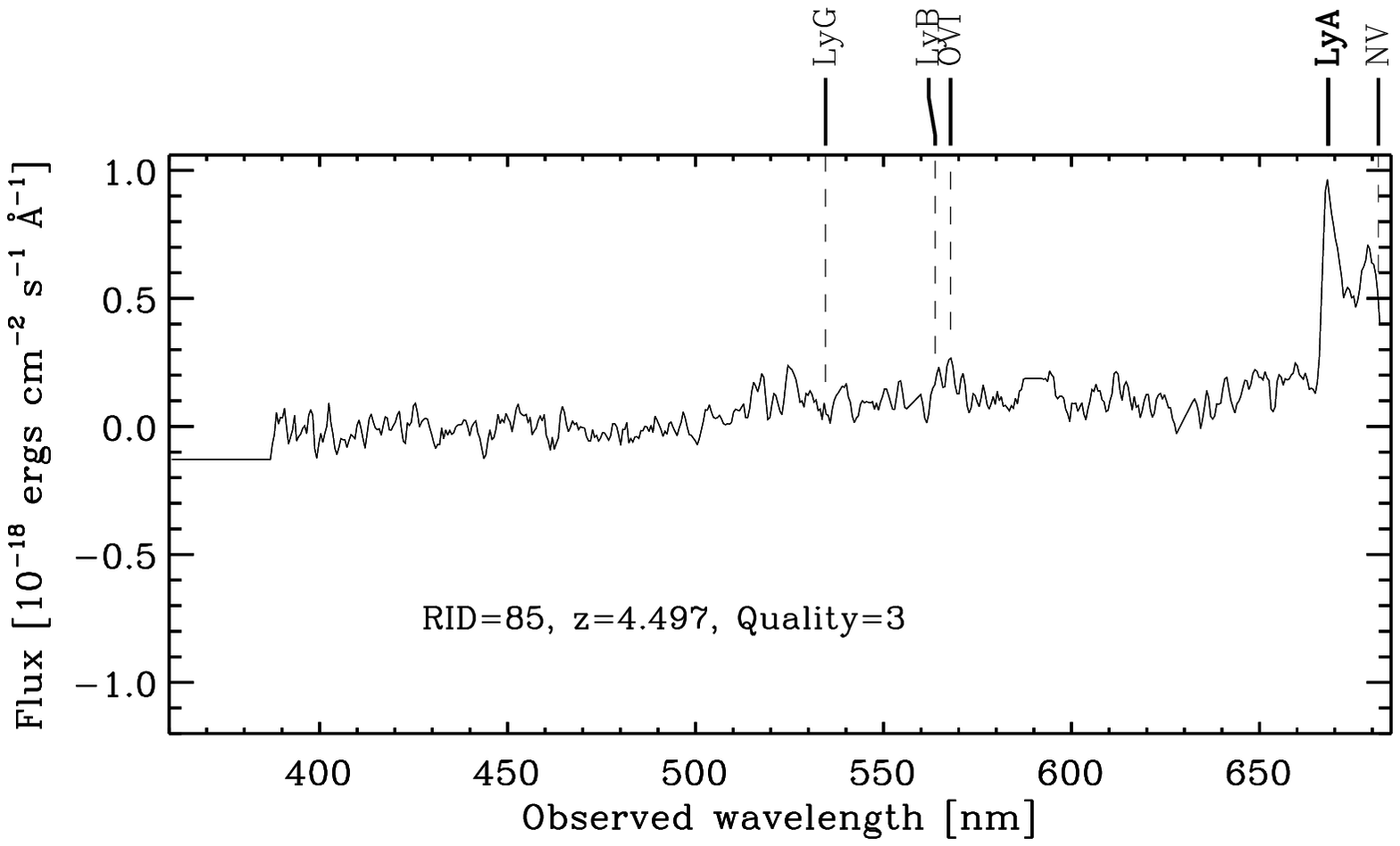}\\
\includegraphics[width=3.in]{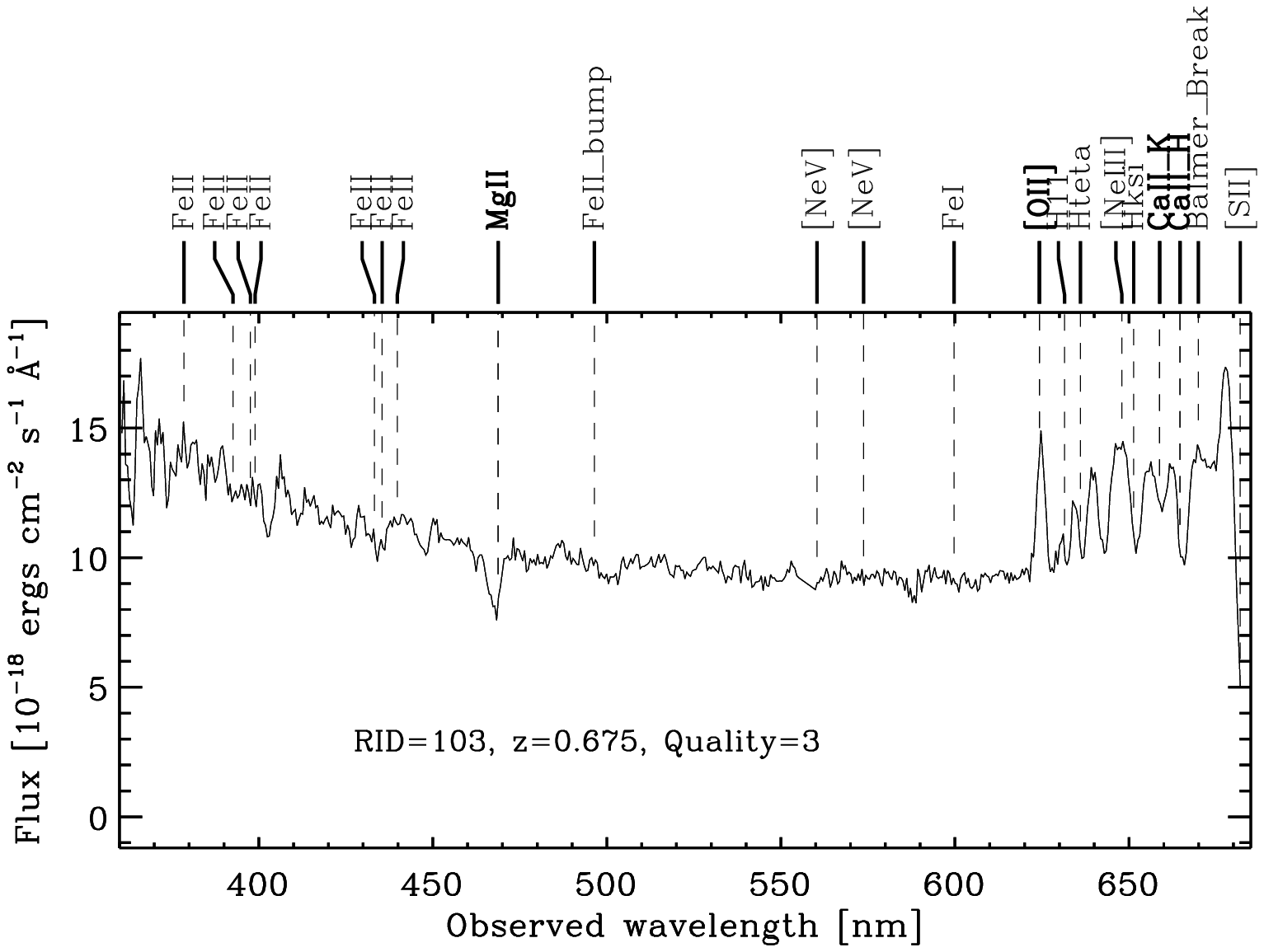}
\includegraphics[width=3.in]{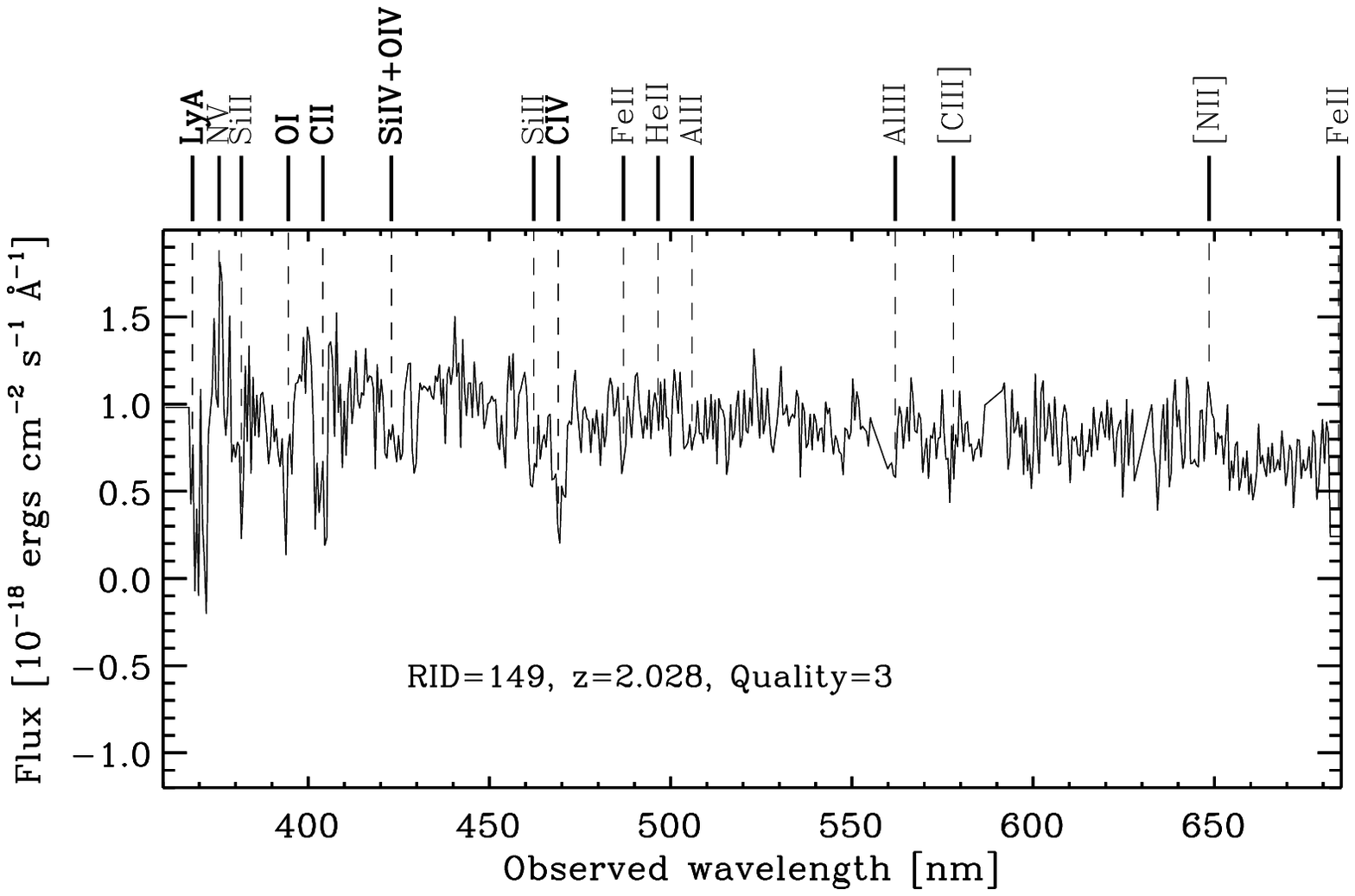}\\
\caption{\small{New VIMOS/VLT spectra of the radio sources.}}
\label{fig_spectra}
\end{center}
\end{figure*}
\clearpage

\addtocounter{figure}{-1}
\begin{figure}
\begin{center}
\includegraphics[width=3.in]{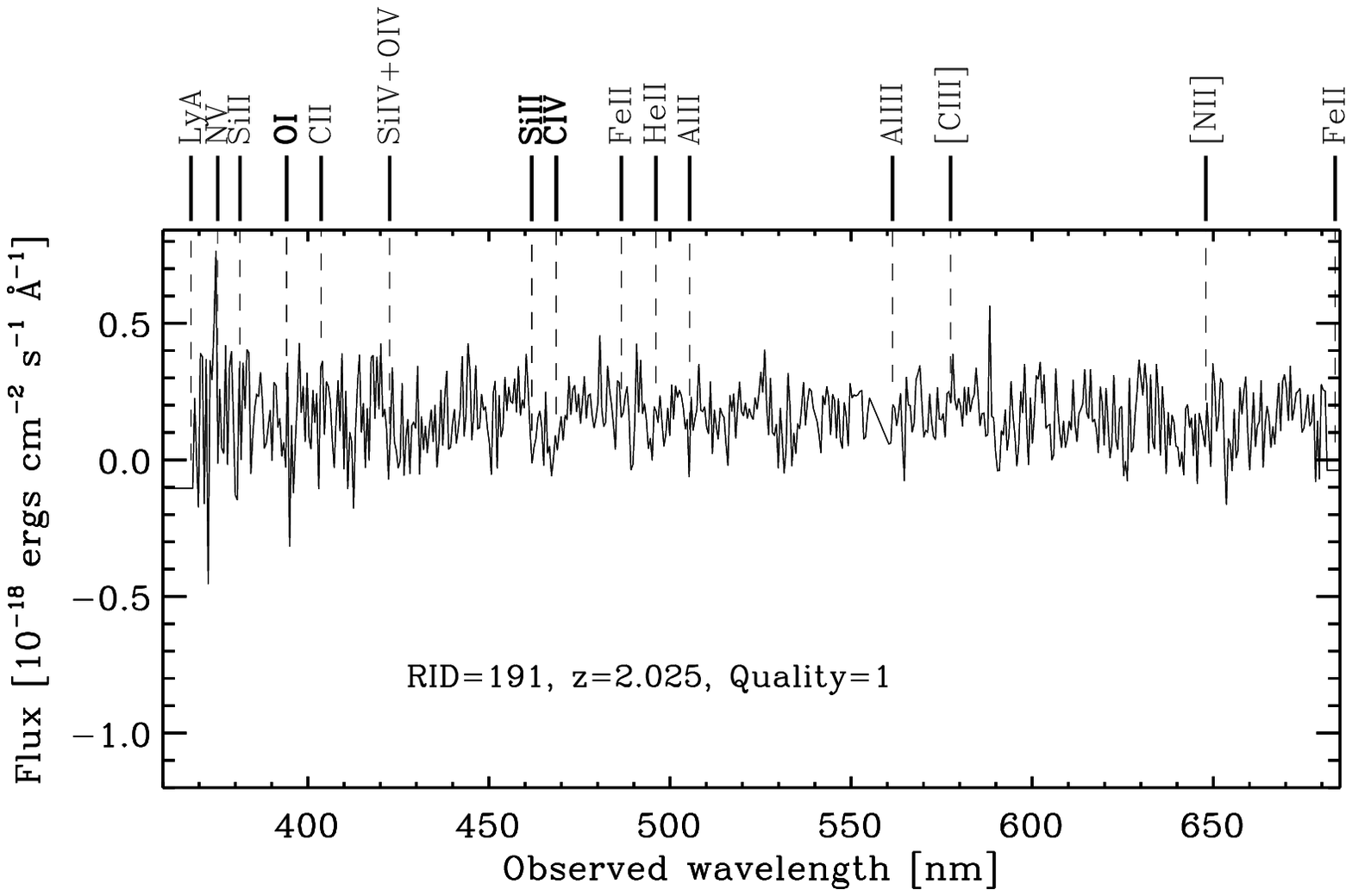}
\includegraphics[width=3.in]{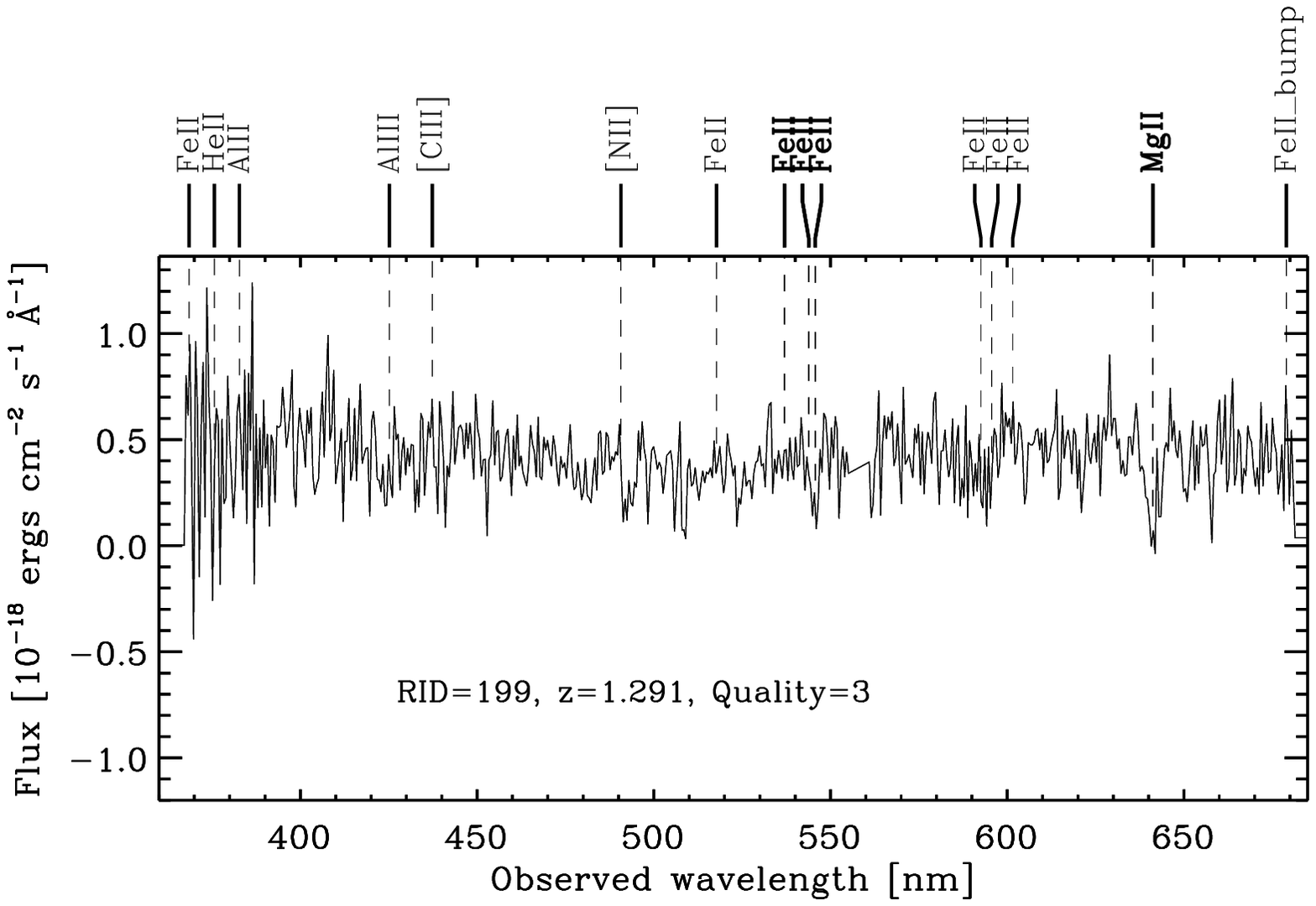}\\
\includegraphics[width=3.in]{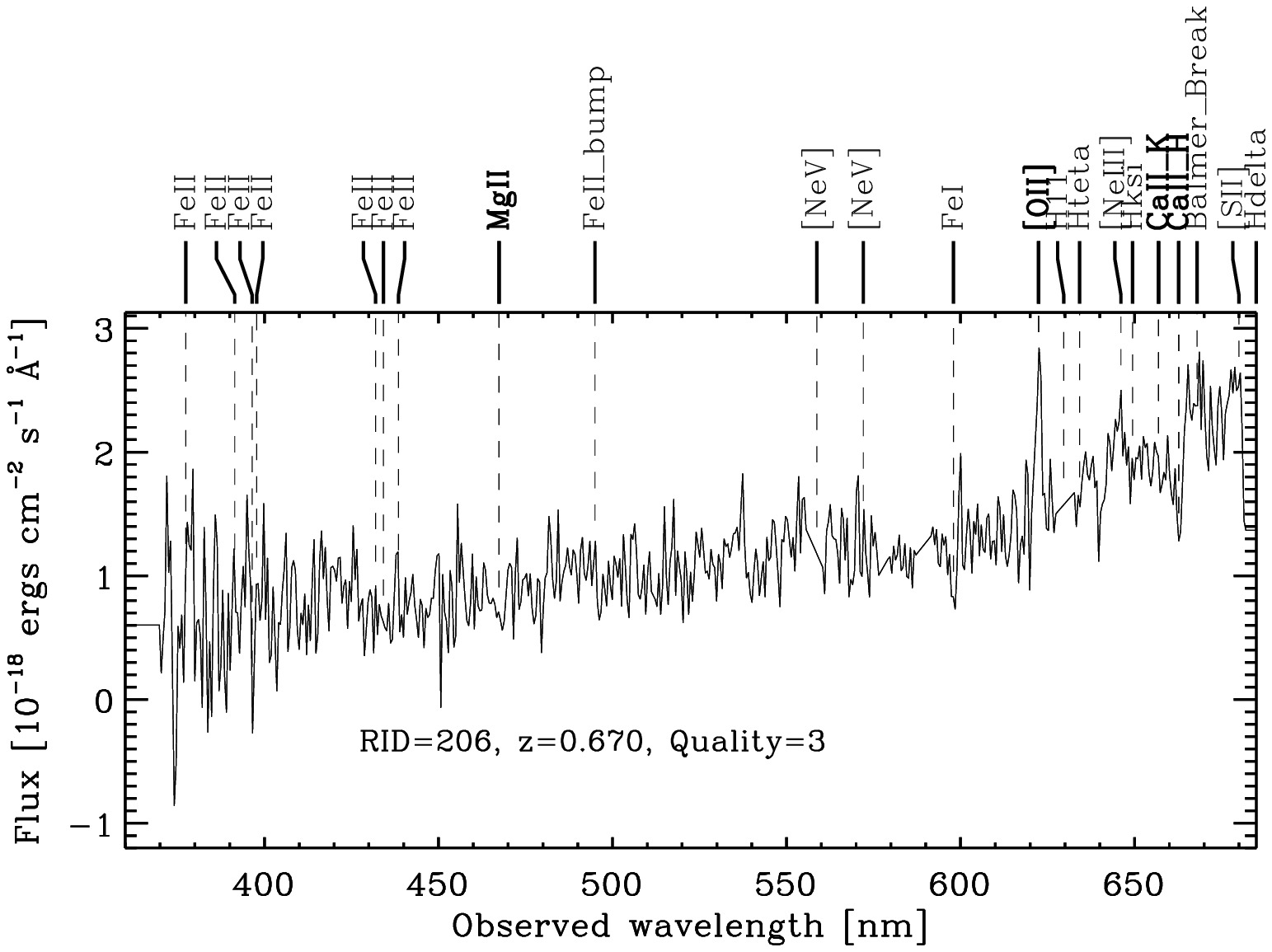}
\includegraphics[width=3.in]{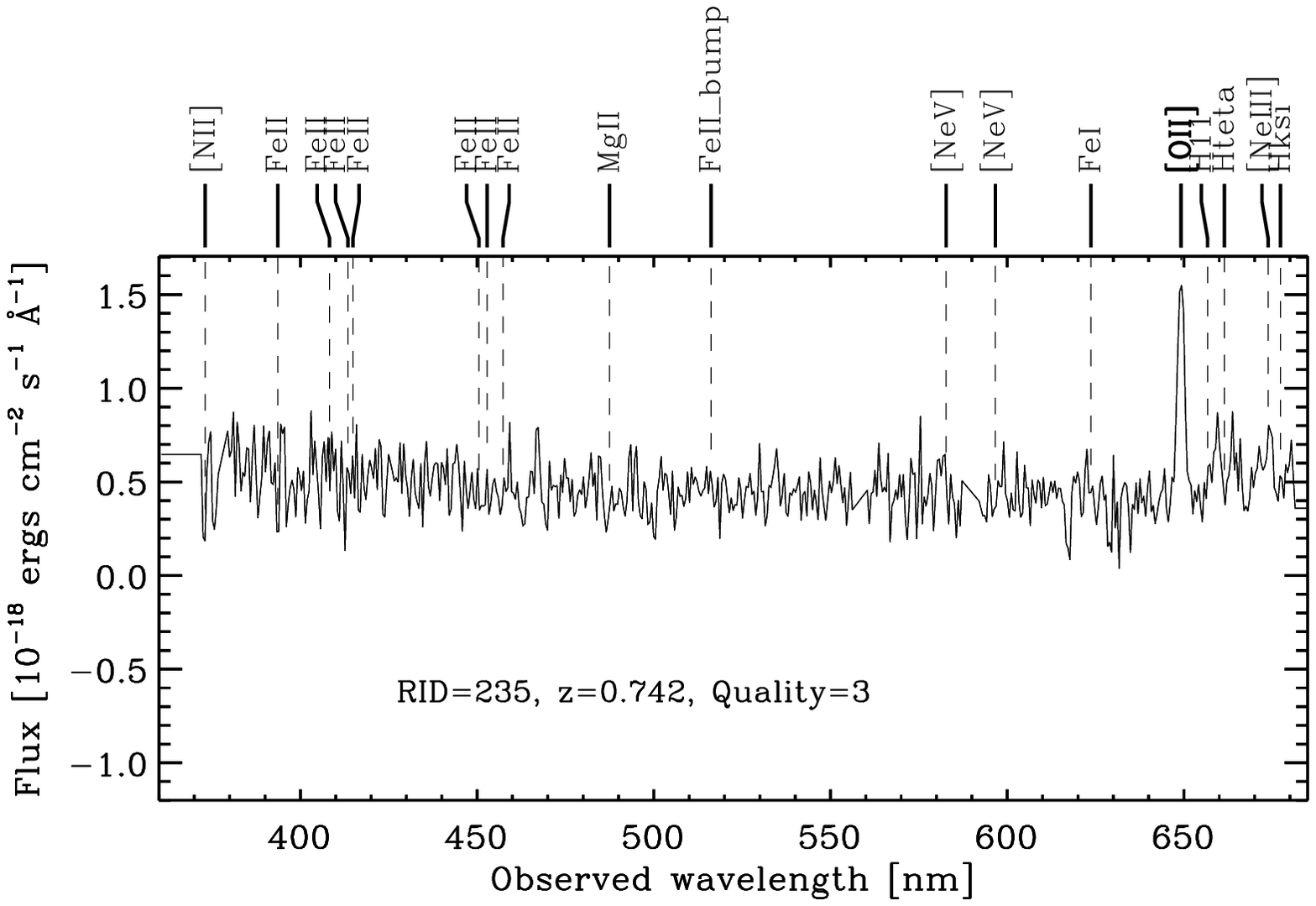}\\
\includegraphics[width=3.in]{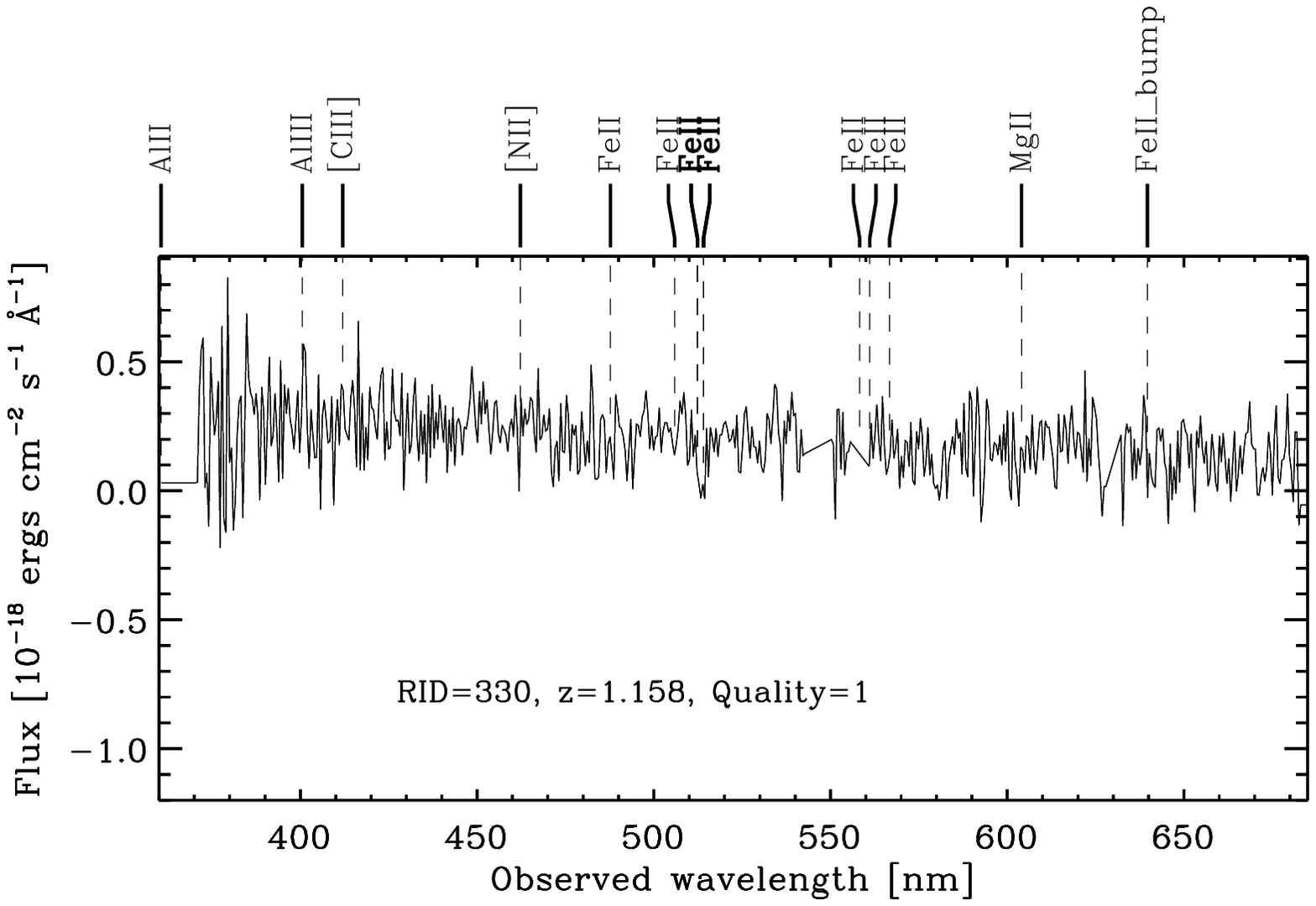}
\includegraphics[width=3.in]{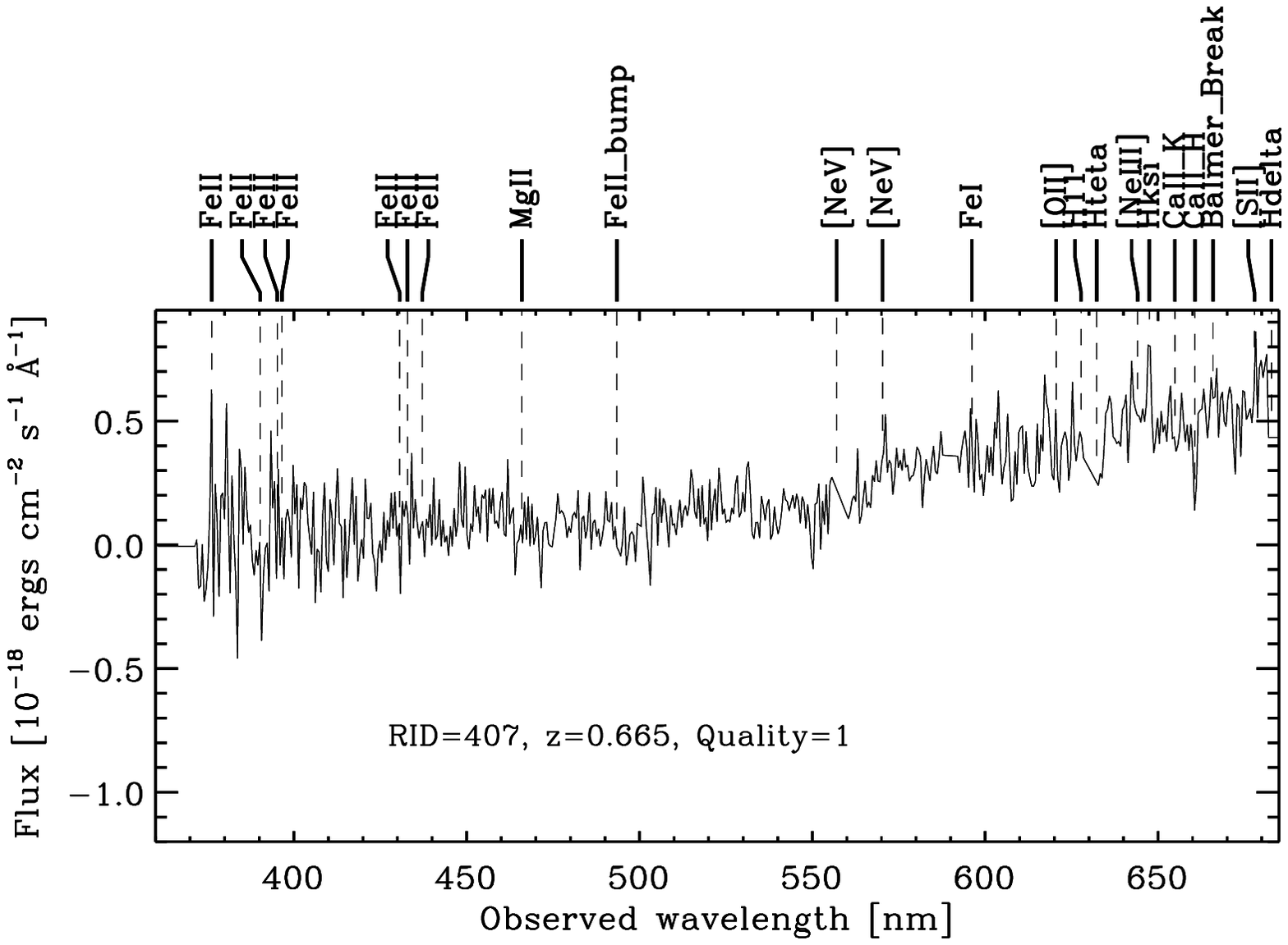}\\
\includegraphics[width=3.in]{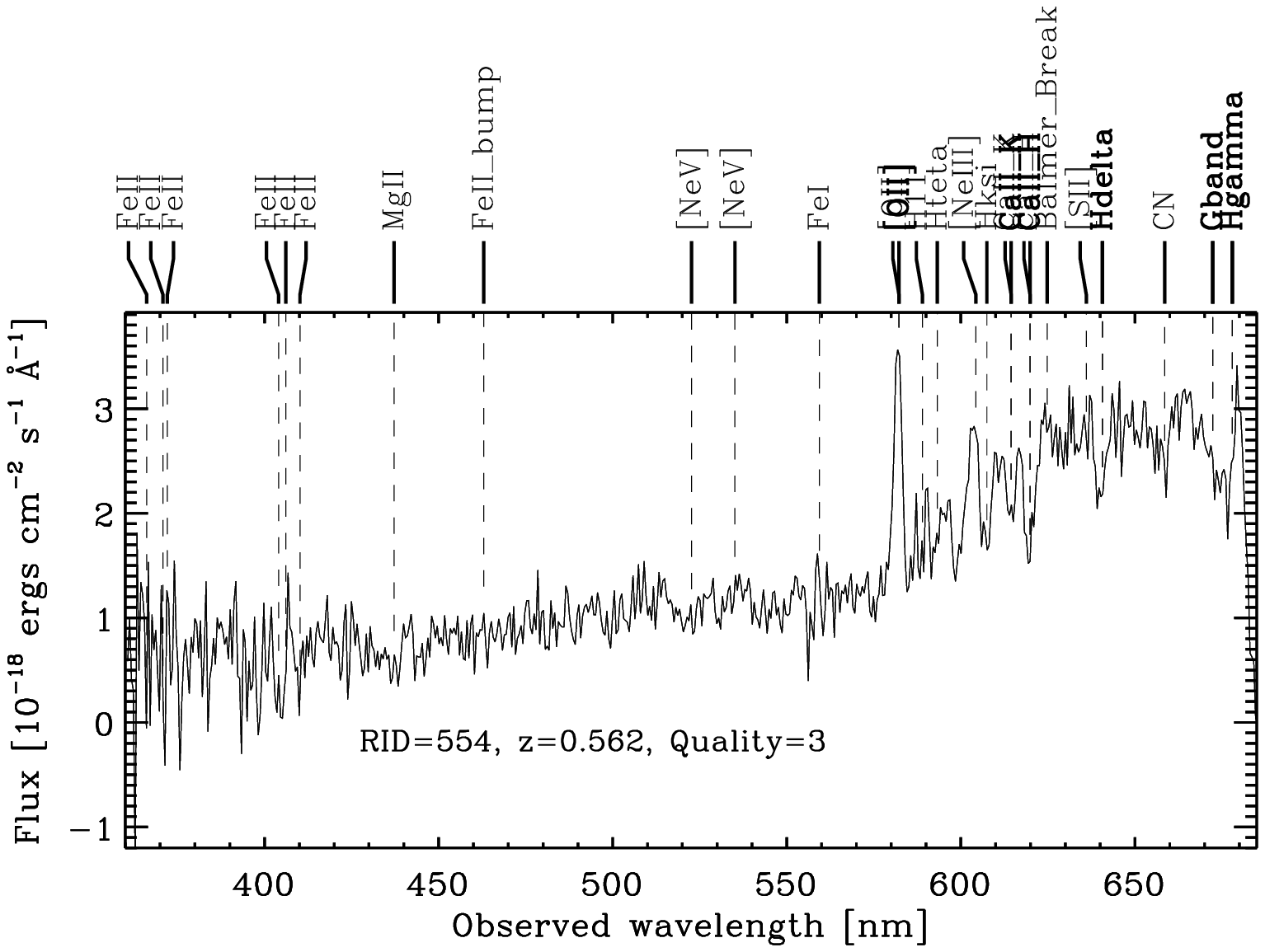}\\
\caption{(Continued.)}
\end{center}
\end{figure}
\clearpage

\section{Notes on individual sources.}
\label{sec-particular}

\begin{itemize}
\item RID 73 (KID 14): there are two possible counterparts for the
  radio lobes but we believe that they are both associated with
  the bright central galaxy. One strong indication for this is the
  strength of the radio flux density, which at 40 mJy is reasonable for a compact double lobed 
  radio galaxy. If separate sources, they would have to both be strong radio
  AGN very close in projection on the plane of the sky.
\item RID 80-85 (KID 18B-18A): they were considered as radio lobes in
  \citet{kellermann08}, but they appear to have two different
  counterparts. Moreover, they have very different flux densities which supports
  the idea that they are not related to the same source. Finally, there is no good candidate for a single radio core.
\item RID 101 (KID 23): double lobed source whose core was
tentatively identified in M08 (KID 23) with a faint (R-band magnitude
$\sim$ 26) galaxy at z=0.999 from the COMBO17 catalog. We think that
this association is very unlikely especially because this possible
counterpart is not detected at any longer wavelength. From the radio
contours we believe that it is a classical radio galaxy. Therefore, we expect for such a galaxy a correlation between the K-band magnitude and the redshift \citep[e.g.,][]{lilly84}. Since we do not detect it in the K-band we think this
source may be a high redshift object. This hypothesis is also supported by the presence of a possible counterpart at 24 micron in the FIDEL catalog. Unfortunately no photometric redshift is available.
\item RID 209 (KID 48): V-shape radio source at redshift 1.3. Given the radio morphology, we considered the possibility that this is a head-tail radio source due to the interaction with the intracluster medium (ICM) in a high-z galaxy cluster. Any cluster, or large group, with ICM density sufficient to
bend the radio jets, would have been clearly detected in the X-ray too. However, we do not observe it in the 4 Ms \textit{Chandra} image. 
Also the redshift distribution of the sources in the region around RID 209 does not show any hint of clustering.  
Therefore, we think is unlikely that the V-shape of this radio source is due to the interaction with the ICM.
There is instead a possible contamination to the flux density of one of the two radio lobes (A) from a superimposed galaxy.
\item RID 283 (KID 73): double-lobe source. There are three optical
  sources in the region of the radio emission but all have reliability under the threshold ($<0.6$) and we believe that none of them are
  associated with the radio source. The counterpart
  of the core is identified with an object detected in the Ks-MUSYC
  catalog.
\item RID 308 (KID 80): bright radio source with possibly one or two
  lobes. However, the quality of the radio image in this region is not good.
\item RID 360 (KID 97): powerful single
component radio source that was not identified in M08. It has a radio flux density of 1.38 mJy. Although we use deeper catalogs, still we are
not able to identify it in any band. The cutouts of this source are shown in Fig. \ref{fc_examples} and they are all blank field. The 5$\sigma$ detection limit for each band is given in Table \ref{tab_aux_cat}. Moreover, this source is in the region covered by the 4Ms \textit{Chandra} observation but it has no X-ray detection. We can therefore put an upper limit on its X-ray flux of $9.1\times10^{-18}$ and $5.5\times10^{-17}$ ergs cm$^{-2}$ s$^{-1}$ 
for the soft and hard band, respectively. 
\item RID 403-406-410-412 (KID 114): this group of sources was at
  first interpreted as a tailed radio source (see radio contours). But
  since we find a clear counterpart for three of them, we consider
  these sources as independent. Source 410 is unidentified.
\item RID 407 (KID 113): bright and extended double lobed source. Close
  to the core position there are many optical sources. In particular
  there is a 21 K-band magnitude galaxy 0.5 arcsec away from the
  expected core position that was automatically selected by our
  method. We consider this association spurious since it would imply
  that this source is far from the K-z relation for radio
  galaxies \citep[e.g][]{lilly84,debreuck02}. Therefore, we manually corrected the identification by
  associating this radio source to a bright elliptical galaxy  (K-band mag=18.4) 2 arcsec away from the centroid of the radio image.  
\item RID 500 (KID 148): this is a complex radio source. We
  identify a clear core and two radio lobes (KID 148A and 148B). There
  are two other components, KID 148D and 148E, possibly associated
  with this source. We found a secure counterpart for 148D and so we
  listed it as a separated source (RID 504). For the 148E component the only counterpart candidate is a faint
  galaxy at $\alpha=$03:32:32.59, $\delta=$-28:03:15.4 with R-mag=23.7, but it is under our reliability threshold and therefore it probably remains unidentified.\item RID 848 (KID 260): the radio source is split into two components both
  associated with the same spiral galaxy.
\end{itemize}




\end{document}